\documentclass[twocolumn,usenatbib]{aastex61}
\usepackage{natbib}

\newcommand{\cosmo}{$H_0 = 67.8$~km~s$^{-1}$~Mpc$^{-1}$, $\Omega_M = 0.308$ and $\Omega_{\Lambda} = 0.692$ \citep{PlanckCollaboration16}}
\newcommand{\Eiso}{$E_{\gamma, \rm iso}$}

\received{}
\revised{}
\accepted{\today}

\submitjournal{ApJ}

\shorttitle{A Search for NS-BH Mergers in the SGRB Population}
\shortauthors{Gompertz et al.}

\begin{document}

\title{A Search for Neutron Star--Black Hole Binary Mergers in the Short Gamma-Ray Burst Population}
\correspondingauthor{Benjamin Gompertz}
\email{b.gompertz@warwick.ac.uk}

\author{B. P. Gompertz}
\affil{Department of Physics, University of Warwick, Coventry, CV4 7AL, UK}

\author{A. J. Levan}
\affil{Department of Physics, University of Warwick, Coventry, CV4 7AL, UK}
\affiliation{Department of Astrophysics/IMAPP, Radboud University, PO Box 9010, NL-6500 GL Nijmegen, The Netherlands}

\author{N. R. Tanvir}
\affiliation{Department of Physics and Astronomy, University of Leicester, Leicester, LE1 7RH, UK}

\begin{abstract}
Short gamma-ray bursts (SGRBs) are now known to be the product of the merger of two compact objects. However, two possible formation channels exist: neutron star  -- neutron star (NS -- NS) or NS -- black hole (BH). The landmark SGRB\,170817A provided evidence for the NS -- NS channel, thanks to analysis of its gravitational wave signal. We investigate the complete population of SGRBs with an associated redshift (39 events), and search for any divisions that may indicate that a NS -- BH formation channel also contributes. Though no conclusive dichotomy is found, we find several lines of evidence that tentatively support the hypothesis that SGRBs with extended emission (EE; 7 events) constitute the missing  merger population: they are unique in the large energy band-sensitivity of their durations, and have statistically distinct energies and host galaxy offsets when compared to regular (non-EE) SGRBs. If this is borne out via future gravitational wave detections it will conclusively disprove the magnetar model for SGRBs. Furthermore, we identify the first statistically significant anti-correlation between the offsets of SGRBs from their host galaxies and their prompt emission energies.
\end{abstract}


\section{Introduction}

Short Gamma-Ray Bursts (SGRBs) are brief, intense flashes of $\gamma$-ray emission, distinct from Long GRBs (LGRBs) in both duration and spectral hardness \citep{Kouveliotou93}. GRB durations are measured by the parameter $t_{90}$; the time in which the central 90 per cent (i.e. 5 -- 95 per cent) of their $\gamma$-ray fluence is detected, and SGRBs typically have $t_{90} < 2$~s. Like LGRBs, their emission is generally well modelled as an explosive event that deposits energy into a collimated highly relativistic jet. The $\gamma$-rays may be produced by interactions between expanding shells of ejecta in this jet \citep{Paczynski86,Rees92}, or by the dissipation of magnetic fields \citep{Usov94}. As the jet expands into the circumstellar medium, it is decelerated through interactions with the ambient environment, and consequently forms a shock front that radiates a synchrotron `afterglow' \citep{Blandford76}. Broadband afterglows from SGRBs are detected across the electromagnetic (EM) spectrum; at X-ray, ultra-violet (UV), optical, infra-red (IR) and (infrequently) at radio frequencies \citep[e.g.][]{Fong15}.

Several lines of evidence now point to the mergers of binary compact objects as the progenitors of SGRBs. While short GRB afterglows were not identified until 2005, several years
after LGRBs \citep{Gehrels05,Hjorth05}, it was immediately apparent that they were spawned from a different population. First, early observations of SGRBs showed no
evidence for any supernova signature (unlike in LGRBs, which are firmly established as core collapse events \citealt{Hjorth03,Levan16}), where one should have been readily visible given the redshift of the bursts and the depth of the observations \citep{Hjorth05b,Rowlinson10}. Second, the host galaxies apparently included ancient elliptical hosts with little or no evidence of star formation
\citep{Gehrels05,Berger05b,Bloom06}, and large samples of host galaxies now show strong support for very different hosts \citep{Fong13}. Furthermore, the bursts are scattered on their host galaxies \citep{Fong13b} and sometimes have no identifiable host at all \citep{Berger10,Tunnicliffe14}. All of this is consistent with a progenitor which can be old, and which has a significant velocity with respect to its host. These combined requirements are entirely consistent with the expectations of compact object mergers whose merger time scales as the fourth power of the initial separation \citep{Peters64}, and which receive significant kick velocities due to mass loss and supernova natal kicks \citep{Belczynski06,Church11}. 

More direct support for a binary merger progenitor was provided with the detection of an infra-red excess observed alongside SGRB 130603B \citep{Tanvir13,Berger13}. This excess is consistent with a radioactive `kilonova' \citep[KN;][]{Li98,Rosswog05,Metzger10b,Barnes13,Metzger17}, in which unstable heavy elements form via rapid neutron capture \citep[$r$-process;][]{Lattimer74,Eichler89,Freiburghaus99} nucleosynthesis and subsequently decay radioactively. The complex electron shells result in large opacities for optical light, and yield a longer-lived infrared transient. Following this event, further KN candidates were proposed after a re-analysis of the archival data of SGRBs 060614 \citep{Yang15}, 050709 \citep{Jin16} and 070809 \citep{Jin20}, and in SGRBs 150101B \citep{Gompertz18,Troja18} and 160821B \citep{Kasliwal17,Jin18,Lamb19b,Troja19}.

In August 2017, the Gamma-ray Burst Monitor \citep[GBM;][]{Meegan09} on board the \emph{Fermi} satellite detected SGRB 170817A. Near-simultaneously, the Advanced Laser Interferometer Gravitational-wave Observatory (aLIGO) and Advanced Virgo gravitational wave (GW) observatories identified a spatially coincident GW signal, GW 170817 \citep{Abbott17b} -- whose chirp was consistent with the merger of two neutron stars (NS). These events triggered a worldwide observing campaign \citep{Abbott17} that subsequently revealed an SGRB \citep{Goldstein17,Savchenko17}, an unpolarized \citep{Covino17} two (or three) component KN \citep{Chornock17,Coulter17a,Cowperthwaite17,Drout17,Evans17,Nicholl17,Pian17,Smartt17,Soares-Santos17,Tanvir17,Villar17} and a rising GRB afterglow \citep{Hallinan17,Margutti17,Margutti18,Troja17b,Troja18b,DAvanzo18,Lyman18,Mooley18b}, suggesting an event that was viewed away from the jet axis \citep{Abbott17b,Haggard17,Kim17,Lamb18,Lazzati18,Mandel18}. The resulting confirmation of the historic first joint GW-EM detection of a merging NS binary \citep{Abbott17b} in the nearby galaxy NGC 4993 \citep{Blanchard17,Hjorth17,Levan17} cemented the link between SGRBs and NS-NS binary mergers. While some models suggest a different source for the $\gamma$-rays in GRB 170817A to more typical cosmological SGRBs \citep[e.g.][]{Kasliwal17a,Gottlieb18,Mooley18}, late-time observations favour a structured jet model \citep{Lyman18,Fong19,Lamb19,Wu19}, likely shared in common with SGRBs as a whole.

These observations mean that we have now firmly identified both binary black hole (BH) and NS-NS mergers. During the recent O3 observing run, the Ligo-Virgo Collaboration (LVC) completed the compact binary merger set with the detection of S190814bv, which was classified as the merger of a neutron star and a black hole \citep[NS-BH;][]{LVCGCNS190814bv}. Because these events contain neutron star material, they are expected to create an electromagnetic counterpart in cases where the neutron star is not swallowed whole by the black hole. Though none has been reported for S190814bv \citep{Dobie19,Gomez19,Andreoni20,Watson20}, the considerably larger distance ($\sim6\times$) compared to GW170817 makes it a much more challenging target. Indeed, simulations suggest that a larger fraction of the neutron star mass may remain outside the black hole in some cases, since the neutron star is ``gradually" disrupted over the course of several periastron passages \citep{Rosswog05,Davies05,Hotokezaka13b}. Based on population synthesis calculations, the volumetric rates of NS-BH mergers has been suggested to be comparable to that of NS-NS binaries \citep{Mapelli18,Eldridge19}, although other simulations find much lower rates \citep{Belczynski17}. Up to the end of 2019,  five NS-NS (including GW 170817) and five NS-BH merger event candidates have been identified in GW \citep[although only one, GW190425, has been confirmed;][]{LIGO20}, at a mean distance of 174~Mpc and 366~Mpc, respectively.  The true rate estimates require a full analysis of the gravitational wave observations, allowing for the duty cycle and the various sensitivities as a function of frequency. However, to first order the rate of detection compared to the volume over which they are seen implies that the relative volumetric rate of NS-BH to NS-NS mergers of $\big(\frac{174}{366}\big)^3 = 0.11$, such that in a volumetric sample of significant size we would expect to see the products of both kinds of merger.

In principle, both NS-NS and NS-BH binaries may create short GRBs, and hence we may expect two sub-populations within the short GRBs. Since the merger process in NS-BH systems can consist of a very different range of mass ratio to the NS-NS case, it is quite plausible that the energy budgets and timescales of NS-BH short GRBs could be different to those of NS-NS. Similarly, since the kick processes operating in NS-BH binaries operate on a binary with greater total mass, which has implications both for the average velocity imparted, and for the binary's ability to stay bound following a strong kick \citep{Repetto17}, it is plausible that the physical locations of NS-BH formed SGRBs could also be different. Previously \citet{Troja08} suggested that SGRBs with extended emission \citep[EE;][]{Norris06} may be produced by NS-BH mergers, based on the presence of EE, a smaller average offset from their host galaxies than is found for SGRBs, and an increased rate of optical afterglow detections. However, \citet{Fong13b} found that the host galaxy offsets of this sample were statistically consistent with SGRBs at large. Both studies were based on samples of limited size (17 SGRBs in \citealt{Troja08}, and 22 in \citealt{Fong13b}). Here, we further explore the possible distinctions between NS-BH and NS-NS progenitors for short GRBs by comparison of various different possible indicators of the two different populations. 

This paper is organized as follows: In Section~\ref{sec:sample} we present our sample for analysis, and investigate its properties in Section~\ref{sec:analysis}. Our findings and conclusions are summarised in Section~\ref{sec:conclusions}. We use the convention $F \propto t^{-\alpha}\nu^{-\beta}$ and a cosmology of \cosmo{} throughout.

\section{Sample}\label{sec:sample}

Our sample consists of GRBs that were classified as `short' in the \emph{Neil Gehrels Swift Observatory} Burst Alert Telescope \citep[BAT;][]{Barthelmy05a} catalog \citep{Lien16} and other published papers \citep{Nysewander09,Fong15}, or by the \emph{Swift}-BAT team in their refined analysis GCN Circulars. Fluences for bursts observed by BAT are from \citet{Lien16}. We also collected data on SGRB 050709 \citep{Villasenor05} from the High Energy Transient Explorer \citep[HETE2;][]{Ricker03}, and on SGRB\,170817A from the GBM GRB catalog \citep{Gruber14,vonKienlin14,Bhat16}.

X-ray data come from the \emph{Swift} X-ray Telescope \citep[XRT;][]{Burrows05}, and were retrieved from the UK \emph{Swift} Science Data Centre \citep[UKSSDC\footnote{\url{http://www.swift.ac.uk}};][]{Evans07,Evans09}. We use the flux density at 1~keV light curves, unless otherwise stated. These are corrected for absorption by multiplying by the counts-to-flux-unabsorbed divided by the counts-to-flux-observed from the late-time photon counting mode spectral fit on the UKSSDC.

Our sample is shown in Table~\ref{tab:sample}, and consists of 39 SGRBs with redshifts. The isotropic equivalent $\gamma$-ray energy (\Eiso) of the prompt emission is calculated from the cataloged fluence using a cosmological k-correction \citep[cf.][]{Bloom01} to account for the shifting rest-frame bandpass when different redshifts are observed. GRB 170817A is known to have been observed off-axis \citep{Abbott17b,Haggard17,Kim17,Lamb18,Lazzati18,Mandel18}, and therefore the measured fluence will be an underestimate. For this reason, it is excluded from energy analyses.

It seems intuitive to also correct $t_{90}$ for the effects of cosmological time dilation using $t_{\rm 90, rest} = t_{90}/(1 + z)$ in order to make temporal comparisons between a sample at varying redshifts. However, \citet{Littlejohns13} showed that the evolution of $t_{90}$ with distance is not so simple; the shifting bandpass results in different measured $t_{\rm 90, rest}$ for the same GRB placed at a different redshift. We therefore discuss both $t_{90}$ and $t_{\rm 90, rest}$ throughout.

\begin{table*}
\begin{center}
\begin{tabular}{cccccccc}
\hline\hline
GRB & $z$ & $t_{90}$ & Fluence & $E_{\gamma ,\rm iso}$ & Burst & Afterglow & Redshift \\
 & & (s) & (erg~cm$^{-2}$) & (erg) & type & type & source \\
\hline
050509B & $0.225$ & $0.02 \pm 0.01$ & $7.13^{+1.26}_{-1.23} \times 10^{-9}$ & $8.07_{-1.39}^{+1.42} \times 10^{47}$ & NC & IF & \citet{Castro-Tirado05} \\
050709$^a$ & $0.16$ & $0.07 \pm 0.01$ & $4.03^{+0.41}_{-0.41} \times 10^{-7}$ & $1.54^{+0.16}_{-0.16} \times 10^{49}$ & NC & & \citet{Fox05} \\
050724$^b$ & $0.257$ & $98.7 \pm 8.56$ & $1.01^{+0.07}_{-0.07} \times 10^{-6}$ & $1.71_{-0.12}^{+0.13} \times 10^{50}$ & EE & ML & \citet{Berger05b} \\
051221A & $0.546$ & $1.39 \pm 0.20$ & $1.16^{+0.02}_{-0.02} \times 10^{-6}$ & $7.27_{-0.13}^{+0.13} \times 10^{50}$ & IXP & ML & \citet{Soderberg06} \\
060502B & $0.287$ & $0.14 \pm 0.05$ & $4.92^{+0.39}_{-0.39} \times 10^{-8}$ & $8.23_{-0.66}^{+0.66} \times 10^{48}$ & NC & & \citet{Bloom07} \\
060614$^c$ & $0.125$ & $109.1 \pm 3.37$ & $1.88^{+0.09}_{-0.08} \times 10^{-5}$ & $7.65_{-0.33}^{+0.36} \times 10^{50}$ & EE & ML & \citet{DellaValle06} \\
060801 & $1.13$ & $0.50 \pm 0.06$ & $8.05^{+0.63}_{-0.63} \times 10^{-8}$ & $9.05_{-0.71}^{+0.71} \times 10^{49}$ & IXP; NC & & \citet{Berger07c} \\
061006$^b$ & $0.438$ & $129.8 \pm 30.7$ & $1.43^{+0.09}_{-0.09} \times 10^{-6}$ & $6.74_{-0.41}^{+0.41} \times 10^{50}$ & EE & IF & \citet{Berger07} \\
061201 & $0.111$ & $0.78 \pm 0.10$ & $3.41^{+0.17}_{-0.17} \times 10^{-7}$ & $9.23_{-0.46}^{+0.46} \times 10^{48}$ & NC & ML & \citet{Stratta07} \\
061210$^b$ & $0.41$ & $85.2 \pm 13.1$ & $1.09^{+0.11}_{-0.11} \times 10^{-6}$ & $4.22_{-0.41}^{+0.42} \times 10^{50}$ & EE & & \citet{Berger07c} \\
061217 & $0.827$ & $0.22 \pm 0.04$ & $4.27^{+0.46}_{-0.45} \times 10^{-8}$ & $3.63_{-0.38}^{+0.39} \times 10^{49}$ & NC & & \citet{Berger07c} \\
070429B & $0.902$ & $0.49 \pm 0.04$ & $6.52^{+0.62}_{-0.61} \times 10^{-7}$ & $1.18_{-0.11}^{+0.11} \times 10^{50}$ & NC & IF & \citet{Cenko08} \\
070714B$^b$ & $0.923$ & $65.6 \pm 9.51$ & $7.39^{+0.57}_{-0.56} \times 10^{-7}$ & $1.16_{-0.09}^{+0.09} \times 10^{51}$ & EE & ML? & \citet{Graham07} \\
070724A & $0.457$ & $0.43 \pm 0.09$ & $3.09^{+0.42}_{-0.40} \times 10^{-8}$ & $1.66_{-0.22}^{+0.23} \times 10^{49}$ & IXP & ML & \citet{Cucchiara07} \\
070729 & $0.8$ & $0.99 \pm 0.17$ & $1.02^{+0.10}_{-0.10} \times 10^{-7}$ & $9.51_{-0.93}^{+0.93} \times 10^{49}$ & NC & & \citet{Berger14} \\
070809 & $0.473$ & $1.28 \pm 0.37$ & $1.02^{+0.09}_{-0.09} \times 10^{-7}$ & $5.42_{-0.47}^{+0.48} \times 10^{49}$ & & ML & \citet{Berger10} \\
071227$^b$ & $0.381$ & $142.5 \pm 48.4$ & $4.94^{+0.76}_{-0.72} \times 10^{-7}$ & $1.96_{-0.29}^{+0.30} \times 10^{50}$ & EE & IF? & \citet{D'Avanzo09} \\
080905A & $0.1218$ & $1.02 \pm 0.08$ & $1.41^{+0.12}_{-0.12} \times 10^{-7}$ & $4.59_{-0.38}^{+0.38} \times 10^{48}$ & IXP; NC & & \citet{Rowlinson10b} \\
090426 & $2.609$ & $1.24 \pm 0.25$ & $1.85^{+0.16}_{-0.16} \times 10^{-7}$ & $2.42_{-0.21}^{+0.21} \times 10^{51}$ & & IF & \citet{Levesque09} \\
090510$^c$ & $0.903$ & $5.66 \pm 1.88$ & $6.17^{+0.55}_{-0.55} \times 10^{-7}$ & $7.66_{-0.68}^{+0.68} \times 10^{50}$ & & ML & \citet{Rau09} \\
090515 & $0.403$ & $0.04 \pm 0.02$ & $2.23^{+0.24}_{-0.24} \times 10^{-8}$ & $7.96_{-0.85}^{+0.85} \times 10^{48}$ & IXP; NC & & \citet{Berger10} \\
100117A & $0.915$ & $0.29 \pm 0.03$ & $9.35^{+0.77}_{-0.77} \times 10^{-8}$ & $1.05_{-0.09}^{+0.09} \times 10^{50}$ & IXP; NC & & \citet{Fong11} \\
100206A & $0.407$ & $0.12 \pm 0.02$ & $1.39^{+0.09}_{-0.09} \times 10^{-7}$ & $3.87_{-0.26}^{+0.26} \times 10^{49}$ & NC & & \citet{Perley12b} \\
100625A & $0.452$ & $0.33 \pm 0.04$ & $2.34^{+0.09}_{-0.09} \times 10^{-7}$ & $8.57_{-0.33}^{+0.34} \times 10^{49}$ & NC & IF & \citet{Fong13} \\
101219A & $0.718$ & $0.83 \pm 0.18$ & $4.34^{+0.15}_{-0.15} \times 10^{-7}$ & $2.96_{-0.10}^{+0.10} \times 10^{50}$ & NC & & \citet{Fong13} \\
111117A & $2.211$ & $0.46 \pm 0.05$ & $1.45^{+0.11}_{-0.11} \times 10^{-7}$ & $3.12_{-0.25}^{+0.25} \times 10^{50}$ & NC & IF & \citet{Selsing18} \\
120804A & $1.3$ & $0.81 \pm 0.08$ & $8.78^{+0.28}_{-0.28} \times 10^{-7}$ & $2.34_{-0.07}^{+0.07} \times 10^{51}$ & & IF & \citet{Berger13b} \\
130603B & $0.356$ & $0.18 \pm 0.02$ & $6.27^{+0.16}_{-0.16} \times 10^{-7}$ & $1.48_{-0.04}^{+0.04} \times 10^{50}$ & NC & ML & \citet{Thoene13} \\
131004A & $0.717$ & $1.54 \pm 0.33$ & $2.76^{+0.12}_{-0.12} \times 10^{-7}$ & $3.55_{-0.16}^{+0.16} \times 10^{50}$ & & IF & \citet{Chornock13} \\
140622A & $0.959$ & $0.13 \pm 0.04$ & $1.32^{+0.23}_{-0.23} \times 10^{-8}$ & $6.98_{-1.20}^{+1.20} \times 10^{49}$ & NC & & \citet{Hartoog14} \\
140903A & $0.351$ & $0.30 \pm 0.03$ & $1.35^{+0.06}_{-0.06} \times 10^{-7}$ & $4.42_{-0.20}^{+0.20} \times 10^{49}$ & NC & ML & \citet{Troja16} \\
141212A & $0.596$ & $0.29 \pm 0.10$ & $7.25^{+0.71}_{-0.71} \times 10^{-8}$ & $5.88_{-0.57}^{+0.58} \times 10^{50}$ & NC & IF & \citet{Chornock14} \\
150101B & $0.134$ & $0.01 \pm 0.01$ & $1.41^{+0.65}_{-0.64} \times 10^{-9}$ & $6.05_{-2.76}^{+2.78} \times 10^{46}$ & NC & & \citet{Levan15b} \\
150120A & $0.46$ & $1.20 \pm 0.15$ & $1.44^{+0.10}_{-0.10} \times 10^{-7}$ & $7.72_{-0.53}^{+0.54} \times 10^{49}$ & IXP & & \citet{Chornock15} \\
150424A$^b$ & $0.3^*$ & $81.1 \pm 17.5$ & $3.29^{+0.34}_{-0.30} \times 10^{-6}$ & $6.25_{-0.57}^{+0.65} \times 10^{50}$ & EE & ML & \citet{Castro-Tirado15} \\
160624A & $0.483$ & $0.19 \pm 0.14$ & $4.30^{+0.55}_{-0.55} \times 10^{-8}$ & $1.42_{-0.18}^{+0.18} \times 10^{49}$ & IXP; NC & & \citet{Cucchiara16} \\
160821B & $0.16$ & $0.48 \pm 0.07$ & $1.15^{+0.07}_{-0.07} \times 10^{-7}$ & $7.41_{-0.47}^{+0.47} \times 10^{48}$ & IXP & ML? & \citet{Levan16b} \\
170428A & $0.454$ & $0.17 \pm 0.03$ & $2.82^{+0.13}_{-0.13} \times 10^{-7}$ & $9.94_{-0.47}^{+0.47} \times 10^{49}$ & NC & IF & \citet{Izzo17} \\
170817A$^{d}$ & $0.0098$ & $2.05 \pm 0.47$ & $2.79^{+0.17}_{-0.17} \times 10^{-7}$ & $9.15^{+0.57}_{-0.57} \times 10^{45}$ & & & \citet{Hjorth17} \\
\hline\hline
\end{tabular}
\caption{\footnotesize{Prompt emission and afterglow characteristics for our sample of SGRBs. Tabulated $t_{90}$ values are in the observer frame. Fluences and energies are in the 15 -- 150~keV BAT bandpass unless otherwise marked. Burst type: NC = `Non-collapsar'; EE = `Extended emission'; IXP = Internal X-ray plateau'. See Section~\ref{sec:subsamples}. The `Afterglow type' column indicates whether each burst falls into the magnetar-like (ML) or injection free (IF) class. Bursts with question marks are not included in the sample analysis in Section~\ref{sec:afterglow}. \emph{a} - $t_{90}$ and fluence (2 -- 400~keV) as measured by HETE, with \Eiso{} k-corrected to 15 -- 150~keV; \emph{b} - EE GRB \citep{Lien16}; \emph{c} - `possible' EE GRB \citep{Lien16}; \emph{d} - This burst is known to have been viewed off-axis. $t_{90}$ and fluence (10 -- 1000~keV) as measured by GBM, with \Eiso{} k-corrected to 15 -- 150~keV. $^*$possibly at $z = 1.0^{+0.3}_{-0.2}$ \citep{Knust17}.}}
\label{tab:sample}
\end{center}
\end{table*}

\subsection{Potential Observable Consequences of Different Progenitors}
\label{sec:expectations}

Since NS-BH and NS-NS mergers may be present in the observed population of SGRBs, it is relevant to consider how they may impact the observable properties of the bursts, their afterglows or host galaxies. 
In particular we may envisage differences in:

\begin{itemize}
\item {\bf Central engine:} An NS-BH merger must result in a post-merger BH central engine, whereas a NS-NS merger may result in a NS remnant (either short-lived or long-lived; see Section~\ref{sec:afterglow}).
\item {\bf Duration:} Simulations \citep[e.g.][]{Rosswog05,Hotokezaka13b} show that NS-BH mergers may disrupt the NS over several passages, with some material ejected to large radii. We therefore might expected NS-BH mergers to be longer (in the rest-frame) and less symmetric than NS-NS events (see Section~\ref{sec:prompt}), although this may also depend sensitively on the mass ratio.
\item {\bf Energetics:} Because of the tidal disruption, the mass accretion onto the BH could yield a greater mass budget and higher energies, although this is likely to depend sensitively on the binary parameters (see Section~\ref{sec:prompt}).

\item {\bf Kilonovae:} Because more matter can be ejected tidally by an NS-BH merger (depending on mass ratio, BH spin, and NS equation of state), we may expect brighter KNe associated with this population
\citep[e.g.][]{Tanaka14,Kawaguchi16}. NS-BH KNe may also be redder due to their ejecta retaining a high neutron fraction and hence being more lanthanide rich \citep[e.g.][see Section~\ref{sec:KNe}]{Metzger17}.

\item {\bf Locations and hosts:} While the formation channels for NS-BH and NS-NS binaries are generally similar, the differences in the masses, and potentially the kicks between NS-BH and NS-NS systems may impart detectable changes in locations. In particular, the higher masses of NS-BH systems would yield shorter merger times (for the same separation and eccentricity), and smaller Blaauw kicks \citep[for the same mass loss;][]{Blaauw61}. Such changes depend sensitively on the precise binary evolution, but may be possible to detect (see Section~\ref{sec:hosts}).
\end{itemize}

It should also be noted that the rates of NS-BH and NS-NS could also be very different. While some simulations suggest rates that are similar
\citep[e.g.][]{Mapelli18,Eldridge19}, others imply much smaller populations of NS-BH binaries \citep[e.g.][]{Belczynski17}, especially those with
mass ratios such that the neutron star is disrupted and not swallowed whole \citep[e.g.][]{Foucart14,Foucart17}. In this case it is possible that the observed population of SGRBs may be dominated by a single channel.  Alternatively, while it is clearly now the case that NS-NS mergers can create an SGRB, it is possible that NS-BH mergers do not. Conversely, given the arguments above, should NS-BH mergers create more energetic GRBs than those from NS-NS systems then they could represent a significant fraction of the observed population (due to Malmquist bias) even if the volumetric rates of NS-BH mergers are much lower. Given the large, poorly-understood uncertainties that go into compact object merger rates, the contribution of NS-BH mergers very much remains an open question.

\subsection{Notable Groupings}\label{sec:subsamples}

At the broadest level, SGRBs were defined by \citet{Kouveliotou93}  as having $t_{90} \leq 2$~s, and by virtue of being spectrally harder than LGRBs. However, 
it is clear that this division does not cleanly split the two populations, and it was based specifically on
the Burst And Transient Source Experiment (BATSE) GRB sample.
Furthermore, there are a few relatively well-known sub-populations of the SGRB class. We outline these below, and track them individually throughout our analysis.

Perhaps the best known SGRB sub-class are those bursts that display a period of softer `extended emission' (EE) after the initial spike \citep{Norris06}, typically lasting several tens, or even hundreds, of seconds. They are distinct from LGRBs (despite $t_{90} \gg 2$~s in some cases) by virtue of their negligible spectral lag between high and low energy photons, which is around $20$ -- $40$ times shorter for SGRBs, and has a distribution close to symmetric about zero \citep{Norris01}. Furthermore, EE GRBs apparently arise from different environments \citep[e.g.][]{Fong13}, and contain significantly less energy than LGRBs \citep[see e.g.][]{Lien16}. 
The EE sub-sample of \emph{Swift} GRBs was originally defined by \citet{Norris10}, and was expanded in \citet{Lien16}. 7/39 bursts in our sample have EE.

EE is defined as a prompt emission phenomenon, but we note that many non-EE SGRBs with $t_{90} < 2$~s have early X-ray emission that occurs on a comparable timescale to the EE ($\sim 100$~s), and must be internal (i.e. non-afterglow) in nature due to the rapid decay at its cessation. GRBs with these internal X-ray plateaus may therefore be related to the EE class, and we highlight them in this context throughout our analysis. 9/39 bursts in the sample show an internal X-ray plateau.

Although $t_{90} \leq 2$~s is commonly used in the literature, the measured duration of a GRB is sensitive to the bandpass of the detector, and it has been shown that the $t_{90}$ threshold for SGRBs is likely to be different depending on which instrument detected it \citep[e.g.][]{Bromberg13,Qin13}. The duration distributions of SGRBs and LGRBs overlap, which means that our sample with $t_{90} \leq 2$~s may still contain some interloping LGRBs. This is well illustrated by GRB 090426, which is nominally an SGRB but may in fact be a collapsar \citep{Antonelli09,Thoene11,Xin11}.

\citet{Bromberg13} sought to identify such interlopers by assigning each burst a probability of being a non-collapsar, $f_{\rm NC}$. This probability reflects the fraction of bursts with a given $t_{90}$ that were found to be non-collapsar (i.e. true SGRBs) according to their fits to the overall GRB duration distribution \citep[see also][]{Bromberg12}. Furthermore, the authors performed their fits on sub-samples of bursts with soft, intermediate and hard spectral indices, resulting in a per-instrument probability that a given burst is a true SGRB based on its $t_{90}$ and spectral hardness.

\citet{Bromberg13} did not publish their best fit parameters for GRBs divided into hard, intermediate, and soft classes, so we obtain $f_{\rm NC}$ values for our sample by interpolating their Table~3. This provides a near-perfect agreement with the calculated values of $f_{\rm NC}$ for SGRBs that featured in their study, indicating our method provides a good approximation of their fits. We adopt a sub-sample of SGRBs for which we find $f_{\rm NC} \geq 0.5$ according to their $t_{90}$s and spectral indices, as measured by \emph{Swift}-BAT. The one exception is GRB 050709, which was discovered by HETE. In this case we used the value of $f_{\rm NC}$ from \citet{Bromberg13}. 22/39 SGRBs in our sample qualify as NC. We have chosen the threshold $f_{\rm NC} \geq 0.5$ so that each individual case is assessed to be more likely a merger event than a collapsar. However, summing the probabilities of being a collapsar ($f_{\rm C} = 1 - f_{\rm NC}$) across our sample indicates that we would expect $2.57$ collapsars to remain. We find that our results are largely insensitive to thresholds above $f_{\rm NC} = 0.5$.

We note that the approach of \citet{Bromberg13} still attempts to assign a probability to a given burst based purely on its duration and spectrum (for which they assume single power-law fits). While this is doubtless a significant improvement over simple duration arguments it also does not capture the complete picture. For example, some bursts in older galaxies or without supernova signatures are assigned high probabilities of being collapsars \citep[e.g. GRB 051221A, 070724;][]{Soderberg06,Kocevski10}. This suggests that while an improvement this approach is
still not definitive. Indeed, further, more complex selection scenarios have also been suggested \citep{Levan07,Zhang09}, although have some risk of confirmation bias in utilizing, e.g. the host galaxy properties, as a diagnostic. As a result, we retain NC GRBs as a sub-sample, rather than accepting them as a definition of SGRBs as a whole.

Any sub-sample designation of the SGRBs in our sample is indicated in Table~\ref{tab:sample}. To summarize, we track extended emission (EE) bursts, which are a known population that show an additional feature in their prompt emission; SGRBs with an internal X-ray plateau, which may be EE bursts where EE is detected only in X-rays; and non-collapsar (NC) bursts, which are a subset of our sample that pass a probabilistic threshold \citep{Bromberg13} of being true SGRBs, instead of interloping LGRBs with $t_{90} < 2$~s. However, we do not pre-suppose in our analysis that any of these observational sub-classes necessarily represent separate progenitors.

\section{Sample Analysis}\label{sec:analysis}

\subsection{Magnetars}\label{sec:afterglow}
Many SGRB X-ray light curves are well fitted by the magnetar model \citep{Zhang01}, in which a rapidly-rotating, highly magnetized NS injects energy into the GRB afterglow via magnetic dipole radiation. This model has been applied to both SGRBs \citep{Fan06,Rowlinson10,Rowlinson13,Fan13,Zhang17} and EE GRBs \citep{Bucciantini12,Gompertz13,Gompertz14,Gibson17,Knust17}. Only NS-NS mergers can produce magnetars, so the presence or absence of these features may possibly be used to identify the two merger types.

To make a robust comparison sample, we apply some restrictions to the fits in order to remove marginal cases. For a magnetar fit to be included, we require that:
\begin{enumerate}
    \item There be more than 5 data points.
    \item The fit features a section of light curve in which the temporal index, $\alpha < 0.75$ (for $F \propto t^{-\alpha}$).
    \item There is no section of the light curve with $\alpha > 2$ after the region fitted with the magnetar model.
\end{enumerate}
The first requirement simply excludes spurious fits. The second ensures that a solution including energy injection is necessary according to the synchrotron closure relations \citep[e.g.][]{Sari98}. This is based on the assumption that the electron energies follow a power law distribution with an index of $p \geq 2$, where $p = 2$ corresponds to a temporal index of $\alpha = 0.75$. The third requirement excludes fits to regions of light curve that are not related to the forward shock afterglow \citep[e.g. the population of light curves attributed to NS collapse to a BH in][]{Rowlinson13}. This avoids selecting all EE and internal X-ray plateau bursts by default, instead focusing on the afterglow light curve.

\begin{figure}
\includegraphics[width=\columnwidth]{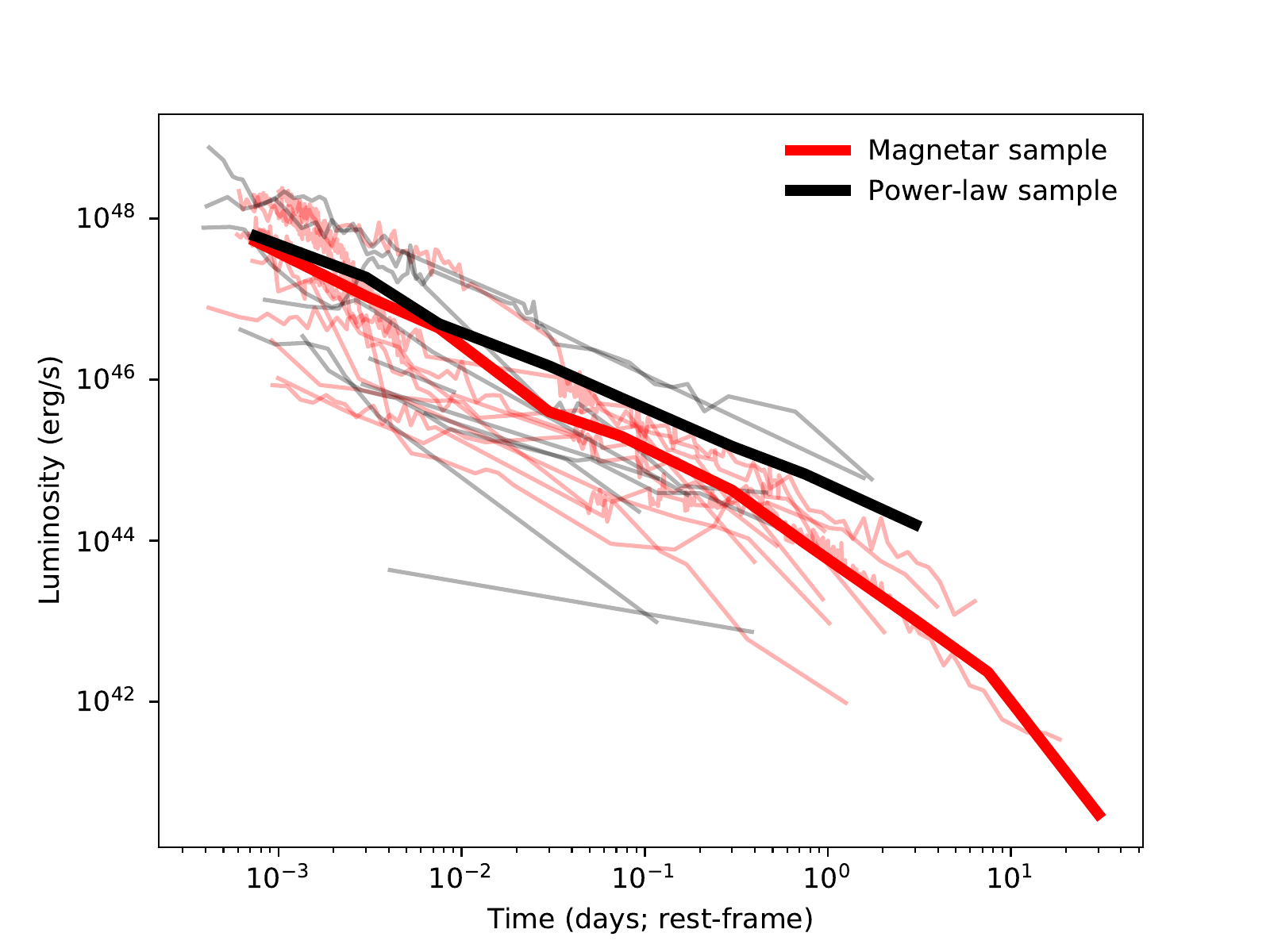}
\caption{The X-ray light curves of bursts with an identified magnetar-like plateau (red), vs those that are well fitted without energy injection  (black). Luminosities for each individual burst are averaged into time bins $0.5$ dex wide, and each time bin is then averaged across the sample to create the mean light curves shown in bold. Luminosities are calculated using a cosmological k-correction to account for the shifting XRT bandbass with redshift.}
\label{fig:xray_comp}
\end{figure}

The comparison sample consists of those SGRB light-curves that were well fitted with a power law (or broken power law) with index values consistent with the expectations for the synchrotron afterglow ($0.75 \leq \alpha \leq 1.5$). We require that this sub-sample have data in the region of $0.1$ -- $1$~days, which is where the energy injection plateaus typically appear. This group are dubbed `injection free' bursts. Table~\ref{tab:sample} lists which category (if any) our sample of SGRBs fall into. The energies and durations in these two sub-samples are not statistically distinct from one another according to either a Kolmogorov-Smirnov (KS) or Anderson-Darling (AD) test, which give a $p_{\rm KS} = 0.79$ ($p_{\rm AD} > 0.25$) and $p_{\rm KS} = 0.42$ ($p_{\rm AD} > 0.25$) chance of both being drawn from the same population for \Eiso{} and t$_{\rm 90}$, respectively.

In order to investigate the relative phenomenology, we construct the mean X-ray light curves of the two afterglow types. We divide the X-ray light curve of each burst into bins that are $0.5$ dex wide in rest frame time. We choose $0.5$ dex as a balance between light curve fidelity and population each bin with data, but the broad trends are insensitive to this choice. In each bin, we find the mean luminosity. The corresponding bins of all GRBs of a given type are then averaged together. Both the mean and individual X-ray light curves of the two sub-samples are shown in Figure~\ref{fig:xray_comp}.

In luminosity space, the features of the average magnetar-like sub-sample light curve become somewhat smoothed out, although a two-plateau morphology is still marginally visible. However, the magnetar-like and injection free sub-samples appear to be equally bright intrinsically. The appearance of magnetar-like features may therefore be influenced by redshift, since they will be easier to detect in more apparently bright bursts. We investigate this first by taking the mean and standard deviation of the redshifts of each sub-sample. For the magnetar-like sub-sample, this is $\bar{z}_{ML} = 0.39 \pm 0.22$, and for the injection free sub-sample it's $\bar{z}_{IF} = 0.99 \pm 0.77$. While these are consistent with one another within errors, it's clear that both the mean and standard deviation for the injection free sub-sample is greater. This is in part due to the fact that it contains two particularly high-z bursts: GRB 090426 \citep[$z = 2.609$;][]{Levesque10e}, and GRB 111117A \citep[$z = 2.211$;][]{Selsing18}. With these excluded, we find $\bar{z}_{IF} = 0.64 \pm 0.32$, which is closer to the magnetar-like sub-sample but nonetheless still higher. A KS test does not find the distribution of redshifts to be statistically distinct ($p = 0.17$), although the AD test does find a $p < 0.05$ chance that the redshifts of the magnetar-like and injection free sub-samples were drawn from the same population. Due to the large number of tests performed, we have collected all of our KS and AD test results together in the Appendix.

Given that the two sub-samples look largely the same in luminosity space, that their $E_{\rm \gamma, iso}$ distributions are not distinct, and that their redshift distributions \emph{are} marginally distinct, the most likely conclusion is that their differing phenomenology in flux space is the result of a selection effect. Features like the energy injection plateau and EE (or the internal X-ray plateau) may simply be harder to identify with a fainter (in the observer frame) and more distant burst. This scenario also explains the higher optical recovery fraction of the magnetar-like sub-sample when compared to the injection free bursts (10/10 and 5/10, respectively). We therefore conclude that the absence of a magnetar-like energy injection plateau is not necessarily an indicator of an NS-BH progenitor population.

\subsection{Prompt Emission}\label{sec:prompt}

\begin{figure}
\includegraphics[width=\columnwidth]{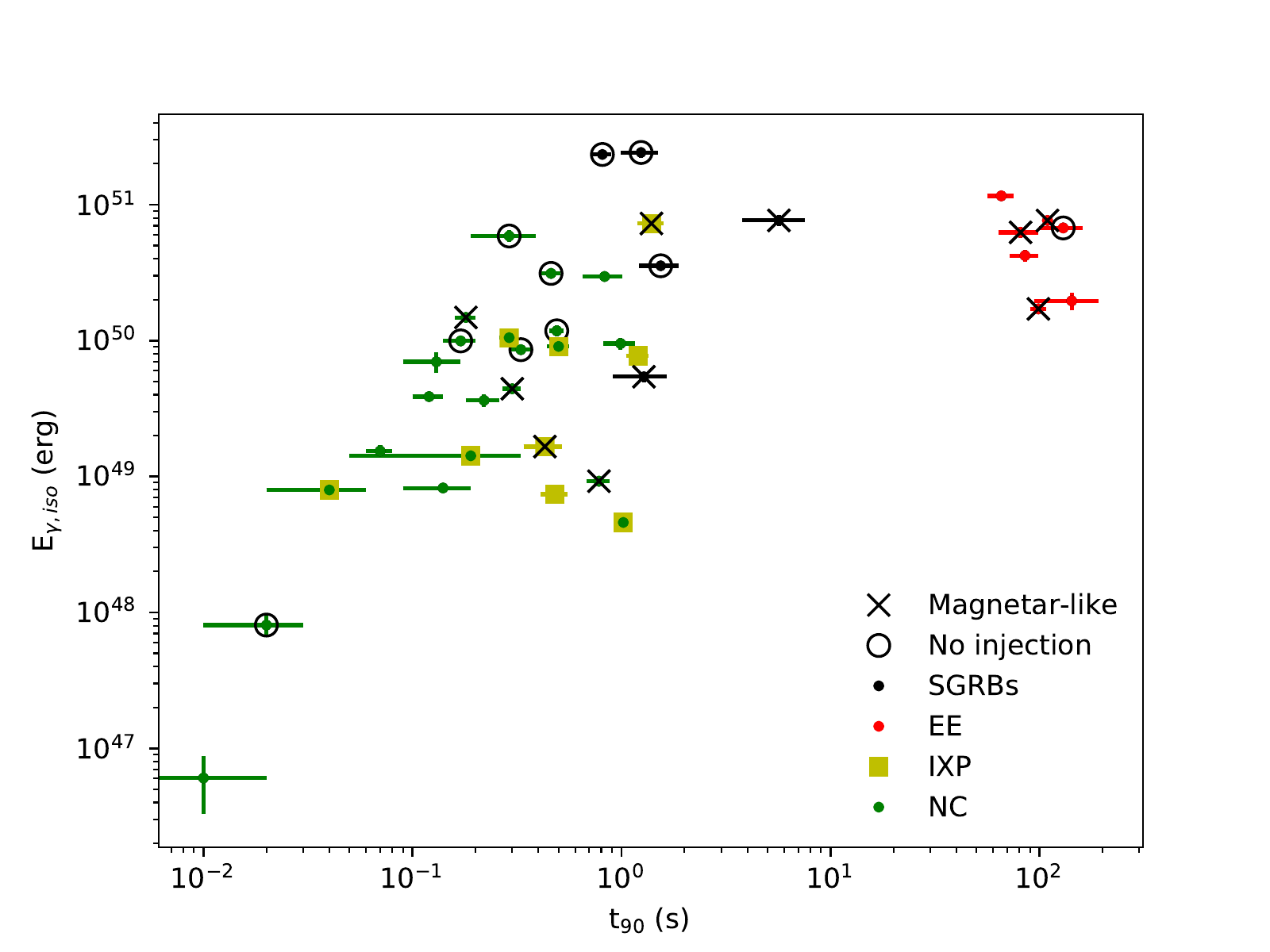}
\caption{The prompt emission energy release, \Eiso{} (15 -- 150~keV), vs $t_{90}$ for our sample. The three sub-samples defined in Section~\ref{sec:subsamples} are marked, as are bursts deemed `magnetar-like' or `injection free' in Section~\ref{sec:afterglow}. The `SGRBs' label indicates bursts that did not fit into any sub-sample.}
\label{fig:Eiso_t90}
\end{figure}

It seems natural to expect that NS-NS and NS-BH binary mergers release different amounts of energy over disparate timescales due to the varying quantity of ejecta produced by mergers of unequal mass binaries \citep[e.g.][]{Davies05,Rosswog05,Hotokezaka13b}. We investigated $t_{90}$ and \Eiso{} in our sample for evidence of a dichotomy. Figure~\ref{fig:Eiso_t90} shows the rest-frame \Eiso{} in the 15 -- 150~keV bandpass vs $t_{90}$ for 38 SGRBs in our sample (170817A is excluded). The $t_{90}$s of the EE bursts are statistically distinct from all other categories, but this group is of course longer by definition. The NC bursts are also statistically distinct from the magnetar-like sub-sample, but this is likely due to the magnetar sub-sample containing many EE bursts. \Eiso{} measurements of the EE bursts are also distinct from the rest of the sample, with a KS (AD) test probability of $p = 1.33 \times 10^{-3}$ ($p = 4.37 \times 10^{-3}$) that they are drawn from the same population. In the lower-left corner lies SGRB 150101B, which has an extremely short duration ($\sim 8$~ms) and anomalously low \Eiso{} ($\sim 6\times 10^{46}$~erg). This burst has been suggested to be an off-axis event like GW/GRB 170817A \citep{Troja18}, although its very short duration is at odds with the comparatively long $t_{90} = 2.05$~s measured for the latter \citep[see also][]{Fong16,Burns18}. No split in either $t_{90}$ or \Eiso{} is apparent in the main body of SGRBs (see the Appendix). The NC sub-sample is statistically distinct in both $t_{90}$ and $E_{\rm iso}$, but primarily as a result of being a large (22/38) sub-sample that sits in opposition of the EE GRBs.

\begin{figure}
\includegraphics[width=\columnwidth]{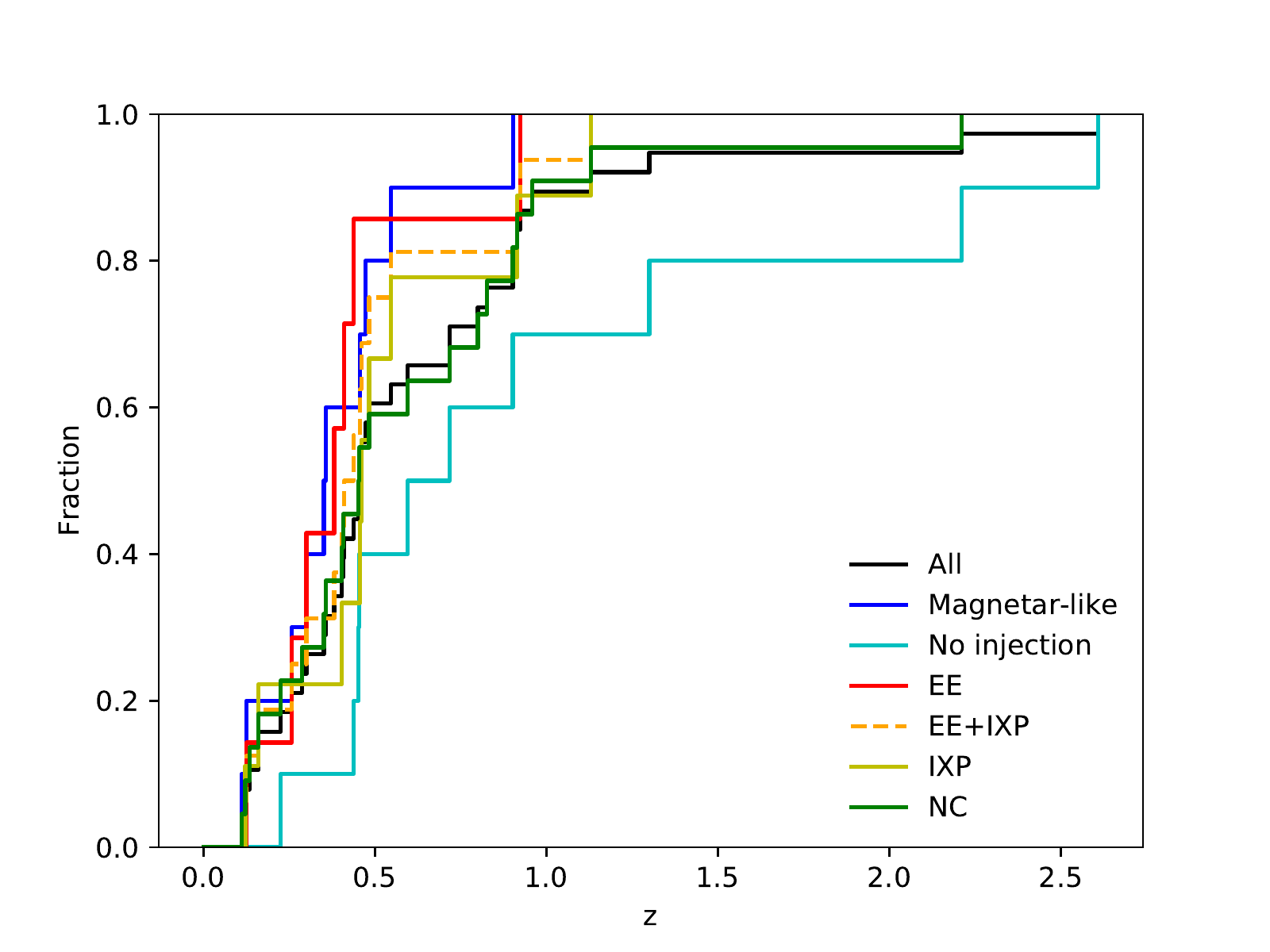}
\caption{The distribution of redshifts within our sample of SGRBs and the sub-samples defined in Section~\ref{sec:subsamples} and Section~\ref{sec:afterglow}. The only statistically significant difference is between the EE GRBs and the injection free GRBs ($p_{\rm KS} = 0.03; p_{\rm AD} = 0.02$).}
\label{fig:z_dist}
\end{figure}

\begin{figure}
\includegraphics[width=\columnwidth]{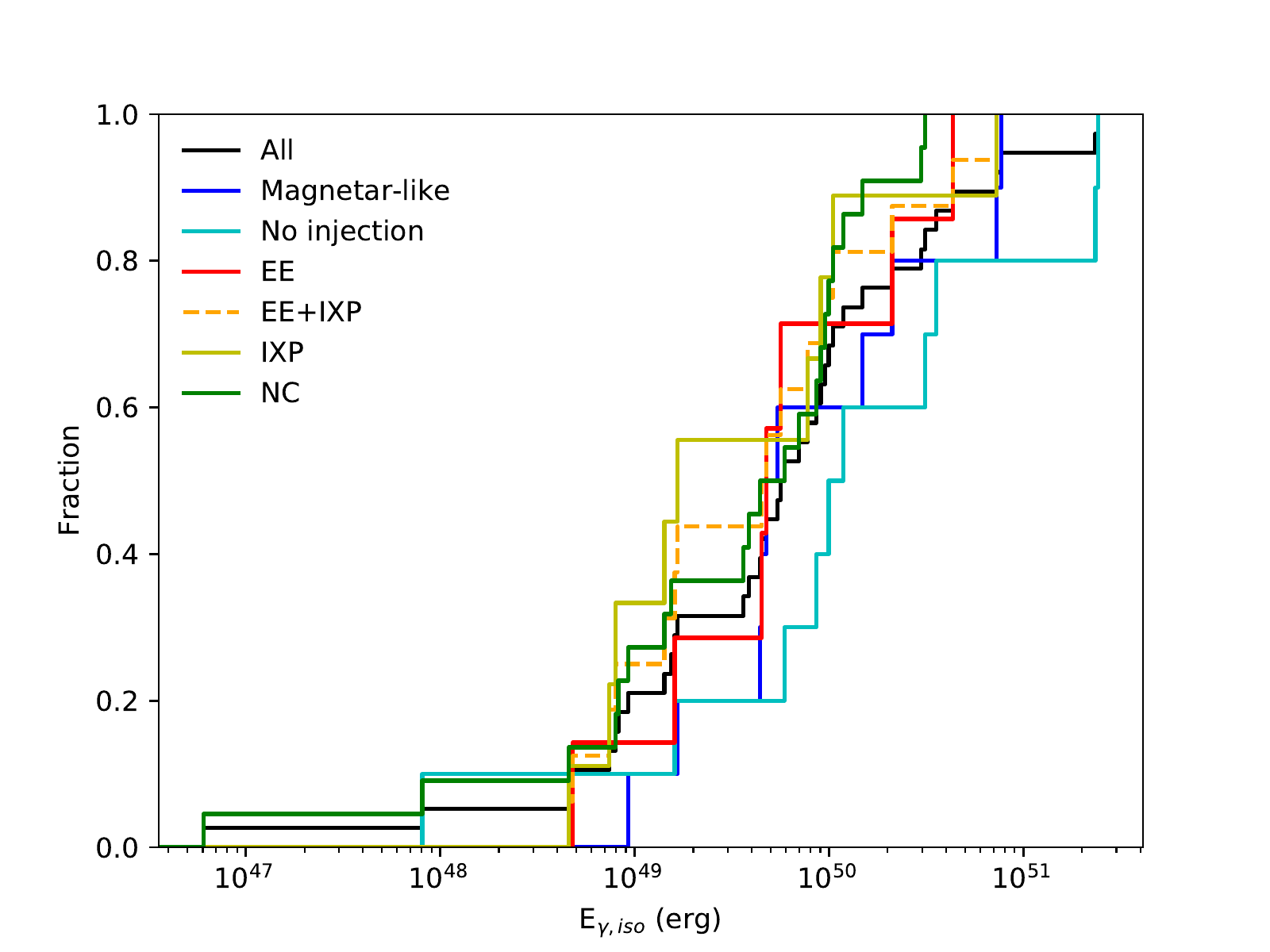}
\caption{The distribution of prompt emission energy releases during the first 2~s for the SGRB sample and the sub-samples defined in Section~\ref{sec:subsamples} and Section~\ref{sec:afterglow}. We also include the EE and internal X-ray plateau sub-samples combined (orange, dashed) to cover the scenario in which internal X-ray plateau bursts are EE bursts that fall short of the BAT bandpass.}
\label{fig:Eiso_2s}
\end{figure}

We also investigate the distribution of redshifts across our full sample and previously defined sub-samples (see the Appendix). Their cumulative distributions in $z$ are shown in Figure~\ref{fig:z_dist}. We find a statistically significant distinction between the redshift distributions of the EE and injection free sub-samples when measured by both the KS ($p = 0.04$) and AD ($p = 0.02$) tests, despite the fact that GRB 061006 is a member of both. The significance of this distinction would be enhanced by excluding this burst, but nevertheless we cannot place too much emphasis on `distinct' populations with overlapping members. However, taking similar but mutually exclusive categories, we find that the injection free sub-sample is also statistically distinct from the magnetar-like sub-sample according to the AD test ($p < 0.05$), but not in the KS test ($p = 0.17$), as previously noted in Section~\ref{sec:afterglow}. The most natural explanation for these distinctions may be that features like EE and injection plateaus are more easily observed at lower redshifts. Some support for this hypothesis may be found in the internal X-ray plateau sub-sample, which are found at slightly higher redshifts than EE GRBs, and whose internal X-ray emission fits the profile of EE at a more extreme redshift. Another possibility is that the separation is due to LGRB interlopers at high redshift in the injection free sub-sample. However, of the two $z > 2$ GRBs, 111117A is also a member of the NC sub-sample, making it a high confidence SGRB, and removing 090426 from the statistical comparison does not invalidate the result. Furthermore, 090426 also exhibited some unusual features: a low neutral hydrogen column density, and time variable Ly$\alpha$ emission, which would be atypical for a long GRB \citep{Thoene11}.
\newline

Based on the prompt emission, the only group of SGRBs to stand out as having a potentially different progenitor is the EE sub-sample, which presents as both higher energy and lower redshift than other SGRBs. We investigate this possibility more thoroughly for the remainder of this Section.

The high \Eiso{} of EE GRBs is in part driven by their long durations. Like regular SGRBs, they feature an initial $\lesssim$ 2~s spike of emission. We extracted the BAT spectra for the first 2~s after trigger for the EE sub-sample and fitted them with a simple power law model to obtain the fluence. We then calculated their \Eiso{}, again using a cosmological k-correction \citep{Bloom01}. The distribution of \Eiso{} of the first 2~s of EE GRBs is compared to the SGRB population at large and the sub-samples defined in Section~\ref{sec:subsamples} and Section~\ref{sec:afterglow} in Figure~\ref{fig:Eiso_2s}. None of the distributions are statistically distinct according to either the KS or AD tests. We note that the distribution of the NC sub-sample \citep[those with $f_{\rm NC} \geq 0.5$;][]{Bromberg13} only deviates from the distribution of the full SGRB sample at the highest energies. This might indicate that our SGRB sample does indeed contain interloping LGRBs.

As a further test of whether EE energies are truly distinct, we compare the energies of our sample in combined $\gamma$- and X-rays out until the mean rest-frame $t_{90}$ of the EE sub-sample, which we find to be $\langle t_{90,\rm rest} \rangle = 75.1$~s. To do this, we took the combined BAT + XRT light curve in the $0.3$ -- $10$~keV bandpass from the UKSSDC, and integrated it out to a time of $t_{\rm int} =  75.1 \times (1+z)$ to obtain the $0.3$ -- $10$~keV fluence.  The flux at $t_{\rm int}$ is obtained by interpolating the flux of the two neighbouring data points. We then converted this fluence to energy. The energies of the SGRBs and EE GRBs calculated this way were found to be identical, with a KS (AD) test showing a $p = 0.39$ ($p = 0.25$) chance that the two were drawn from the same population. This may suggest that an EE phase is present in all (or most) SGRBs \citep[a conclusion that is supported by][]{Kisaka17}, but in most cases falls outside of the $15$ -- $150$ bandpass of BAT. Our sub-sample of internal X-ray plateau bursts may be further evidence of this.

Standardizing the time in which \Eiso{} is determined appears to indicate that EE GRBs are only distinct because they are longer than normal SGRBs. Nonetheless, \citet{Norris10} showed that EE should be detectable if it were present in the majority of SGRBs nominally without EE, based on the ratio ($R_{\rm int}$) of the average EE flux to the peak prompt emission `Initial Pulse Complex' flux across the EE sub-sample. They find a range of $3 \times 10^{-3} \lesssim R_{\rm int} \lesssim 8 \times 10^{-2}$ in the EE sub-sample, but place a $2\sigma$ limit of $R_{\rm int} < 8 \times 10^{-4}$ on this ratio for their sample of an additional 39 SGRBs that do not show EE. Their results suggest that EE is a separate group rather than part of a continuum, and that it is truly absent in 3/4 of SGRBs, rather than just undetected. Conversely, \citet{Perley09a} argue that EE and non-EE bursts may form a continuum in their ratios of prompt spike to EE fluence, with the intermediate values of this ratio populated by BATSE and HETE bursts.
\newline

\begin{figure}
\includegraphics[width=\columnwidth]{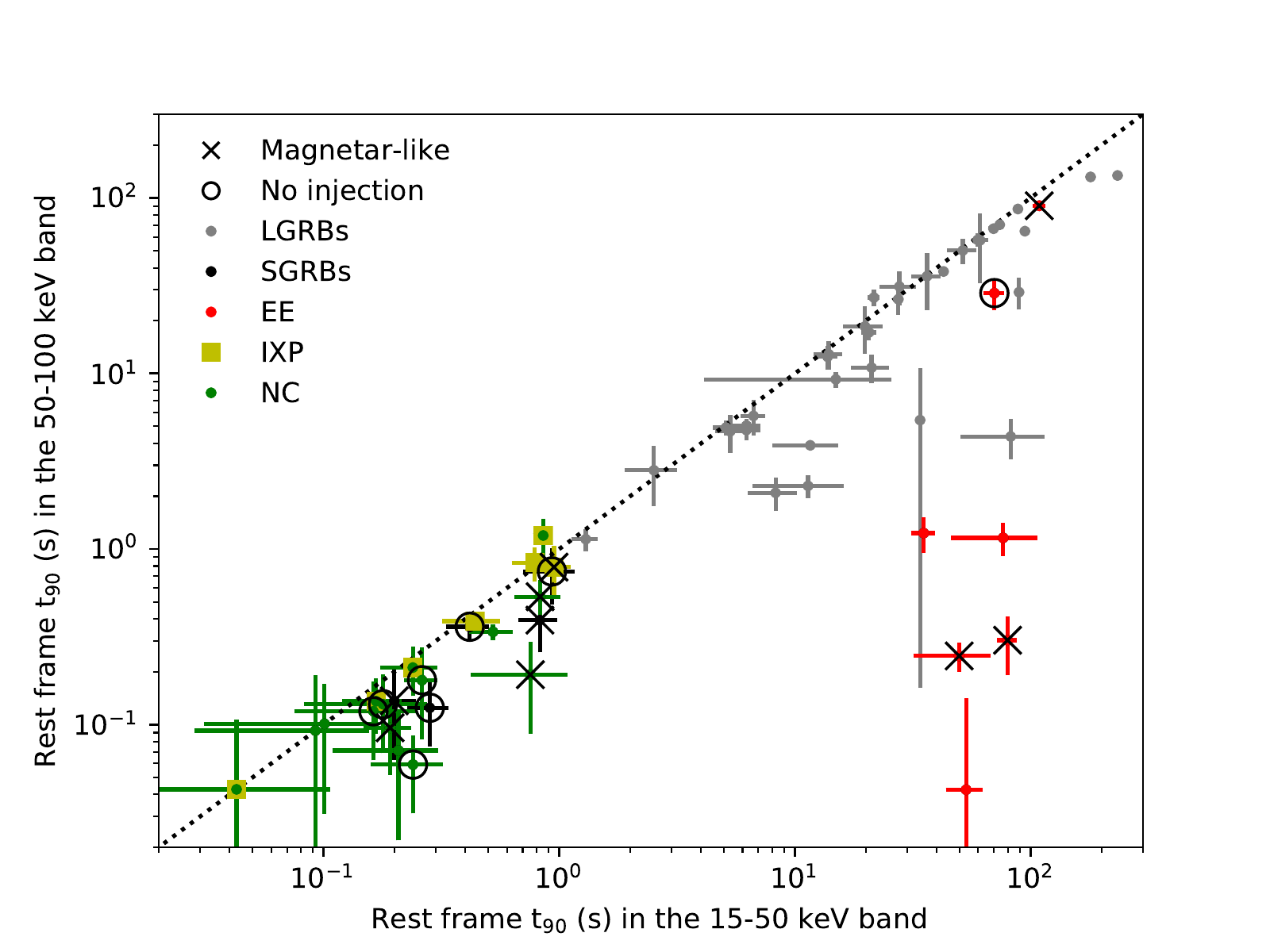}
\caption{Rest-frame $t_{90}$ measured in the 50-100~keV band (y-axis) vs the 15-50~keV band (x-axis). The dotted line marks a 1:1 ratio. Our SGRB sub-samples are compared to a sample of LGRBs from \citet{Gompertz18b}, shown in grey. SGRBs that do not fit into any sub-sample are shown in black. While most bursts track the 1:1 line, the EE sub-sample (red, lower right) yield significantly shorter $t_{90}$ measurements in the higher energy band (with the notable exceptions of 060614 and 061006).}
\label{fig:t90_bands}
\end{figure}

One possible cause of the apparent division between EE and non-EE SGRBs could be the softer EE component falling outside the BAT bandpass in many cases. Figure~\ref{fig:t90_bands} shows the rest-frame $t_{90}$ for BAT light curves that we created in 15 -- 50~keV and 50 -- 100~keV spectral bins\footnote{For GRBs 100117A and 100625A, the data used for duration determination were limited to the first 500s. For 160821B, it was the first 400s. This is due to a sharp rise in background counts in these cases (likely due to the SAA) that causes spurious background subtractions and duration determinations if data beyond these times are included. See: \url{swift.gsfc.nasa.gov/results/batgrbcat/GRB100117A/data\_product/comment.txt};
\url{swift.gsfc.nasa.gov/results/batgrbcat/GRB100625A/data\_product/comment.txt}; \url{swift.gsfc.nasa.gov/results/batgrbcat/GRB160821B/data\_product/comment.txt}}. In the 15 -- 50~keV range, all seven EE GRBs exhibit $t_{90} \sim 100$~s, as they do in the full 15 -- 150~keV bandpass. However, in the 50 -- 100~keV range, only two of the EE bursts continue to present $t_{90} \gtrsim 2$~s (060614 and 061006). This highlights the band-sensitivity of $t_{90}$, and the softness of the EE component compared to the prompt spike. We also present a comparison sample of LGRBs, taken from the sample of \citet{Gompertz18b}, in Figure~\ref{fig:t90_bands}. In general, LGRBs follow fairly closely to the 1:1 ratio of $t_{90}$ between the two chosen energy bands, in common with the short GRBs. This fact reinforces the unusual nature of the EE GRBs. Some LGRBs fall below the 1:1 line, and this could be evidence of cross-contamination between the EE and long samples. Internal X-ray plateau bursts track the 1:1 line, meaning they do not show the duration excess in the softer band like EE bursts do. However, in the $0.3$ -- $10$ keV bandpass of the \emph{Swift} XRT, their measured duration would clearly be longer.

\begin{table*}
\begin{center}
\begin{tabular}{cccccc}
\hline\hline
GRB & log M$_*$ & Host & $r_l$ & $r_h$ & Source \\
 & ($M_{\odot}$) & type & (kpc) & & \\
\hline
050509B & $11.08\pm0.03$ & early & $63.7\pm12.2$ & $3.04\pm0.58$ & [1,2] \\
050709 & $8.66\pm0.07$ & late & $3.64\pm0.027$ & $1.75\pm0.01$ & [1,2] \\
050724 & $10.64\pm0.05$ & early & $2.63\pm0.079$ & $0.49\pm0.01$ & [1,2] \\
051221A & $8.61\pm0.64$ & late & $2.18\pm0.19$ & $0.84\pm0.07$ & [1,2] \\
060502B & $11.8$ & early & $73\pm19$ & & [3] \\
060614 & $7.95\pm0.13$ & & $0.80\pm0.03$ & & [1,4] \\
060801 & $9.1$ & late & $19.7\pm19.8$ & & [5,6] \\
061006 & $10.43\pm0.23$ & late & $1.30\pm0.24$ & $0.40\pm0.07$ & [1,2] \\
061201* & & & $32.47\pm0.06$ & $14.91\pm0.03$ & [7] \\
061210 & $9.6$ & late & $10.7\pm9.7$ & & [5,6] \\
061217* & $9.1$ & late & $55\pm28$ & & [5,6] \\
070429B & $10.4$ & late & $< 11.41$ & $< 2.25$ & [5,7] \\
070714B & $9.4$ & late & $12.21\pm0.87$ & $4.56\pm0.33$ & [5,7] \\
070724A & $10.1$ & late & $5.46\pm0.14$ & $1.50\pm0.04$ & [5,7] \\
070729 & $10.6$ & early & & & [5,8] \\
070809 & $11.4$ & early & $33.22\pm2.71$ & $9.25\pm0.75$ & [5,7] \\
071227 & $10.4$ & late & $15.50\pm0.24$ & $3.28\pm0.05$ & [5,7] \\
080905A & $10.3\pm0.3$ & late & $17.96\pm0.19$ & $10.36\pm0.10$ & [7,8,9] \\
090426 & & late & $0.45\pm0.25$ & $0.29\pm0.14$ & [7,8] \\
090510 & $9.7$ & late & $10.37\pm2.89$ & $1.99\pm0.39$ & [5,7] \\
090515* & $11.2$ & early & $75.03\pm0.15$ & $15.53\pm0.03$ & [5,7] \\
100117A & $10.3$ & early & $1.32\pm0.33$ & $0.57\pm0.13$ & [5,7] \\
100206A & $10.8$ & late & $21.9\pm18.1$ & & [4,8] \\
100625A & $10.3$ & early & $< 19.8$ & & [4,8] \\
101219A* & $9.2$ & late & $< 24.9$ & & [4,8] \\
111117A & $9.9\pm0.2$ & late & $8.5\pm1.7$ & & [10] \\
120804A & $10.8$ & late & $2.2\pm1.2$ & & [8,11] \\
130603B & $9.7$ & late & $5.21\pm0.17$ & $1.05\pm0.04$ & [7,8] \\
140903A & $10.61\pm0.15$ & late & & & [12] \\
150101B & $10.85^{+0.07}_{-0.17}$ & early & $7.35\pm0.07$ & $0.77\pm0.02$ & [13] \\
170817A & $10.65\pm0.03$ & early & $2.125\pm0.001$ & $0.64\pm0.03$ & [14,15] \\
\hline\hline
\end{tabular}
\caption{Host galaxy properties of our sample. The host-normalised radius ($r_h$) column is in units of host effective radii ($r_e$). Where $r_e$ was reported in kpc in the literature, we convert to units of host effective radii by dividing $r_l$/$r_e$ and adding any errors in quadrature. Several bursts in our sample are omitted from the table because they do not have information available in any given category. *given host has a probability of chance coincidence $> 0.05$ \citep{Fong13}.\newline
\emph{References: } [1] - \citet{Savaglio09}; [2] - \citet{Fong10}; [3] - \citet{Bloom07}; [4] - \citet{Li16}; [5] - \citet{Leibler10}; [6] - \citet{Troja08}; [7] - \citet{Fong13b}; [8] - \citet{Berger14}; [9] - \citet{Rowlinson10b}; [10] - \citet{Selsing18}; [11] - \citet{Berger13b}; [12] - \citet{Troja16}; [13] - \citet{Fong16}; [14] - \citet{Blanchard17}; [15] - \citet{Levan17}}
\label{tab:hosts}
\end{center}
\end{table*}

\subsection{Kilonovae}\label{sec:KNe}

The luminosity and rate of evolution of KN signatures depend on several parameters, including the mass ejected, and the velocity and opacity of the ejecta \citep[e.g.][]{Barnes13,Hotokezaka13b,Kawaguchi16,Metzger17,Tanaka18,Barbieri19,Kawaguchi20}. The higher binary mass ratio implicit in NS-BH mergers when compared to NS-NS mergers may therefore result in a larger dynamical ejecta mass during the merger process. The result of this would be a KN with a brighter infrared component \citep[e.g.][]{Metzger17,Kawaguchi20}, although the emission may be indistinguishable from NS-NS mergers for a lower mass ($< 5$M$_{\odot}$) BH  \citep[e.g.][]{Foucart19}. We therefore investigated the population of known (or suspected) KNe for any bimodality.

There appears to be a significant diversity of KN emission in the SGRB sample \citep{Kasliwal17,Fong17,Gompertz18,Ascenzi19}; cosmological KN candidates are typically brighter than AT2017gfo, but several bursts with constraining deep limits do not exhibit any KN emission at all. There are currently six SGRBs that contain KN candidates: 050709 \citep{Jin16}; 060614
\citep{Yang15}; 070809 \citep{Jin20}; 130603B \citep{Tanvir13,Berger13}; 150101B \citep{Gompertz18,Troja18} and 160821B \citep{Kasliwal17, Jin18,Lamb19b,Troja19}, as well as AT2017gfo; the KN associated with GRB 170817A \citep{Chornock17,Coulter17a,Cowperthwaite17,Drout17,Evans17,Nicholl17,Pian17,Smartt17,Soares-Santos17,Tanvir17,Villar17}. In addition, \citet{Gompertz18} identified a further three SGRBs with deep limits that are constraining to an AT2017gfo-like KN: GRBs 050509B, 061201 and 080905A.

Direct comparisons between SGRB KN candidates are extremely difficult to make because the data are often sparse and the magnitudes contain a varying degree of contamination from the GRB afterglow. The available filters are also inconsistent. Variations within the KN behaviours themselves are therefore extremely difficult to separate from observational uncertainties. Parameters such as the ejecta mass and ejecta velocity are highly dependent on the model the data are fitted to. Furthermore, accurately estimating the blackbody temperature requires better SED coverage than is typically available - and would best be done at a consistent time in order for meaningful comparisons to be made anyway.

We attempt to compare our KN candidates in three different ways:
\begin{enumerate}
    \item By comparing the magnitudes of the SGRB KN candidates relative to the AT2017gfo KN model in either the i or r filters at 1 -- 3 days post-merger \citep[following the method of][]{Gompertz18}.
    \item By applying an approximate multiplication factor to the AT2017gfo models after transposition to the redshift of the SGRB KN candidate so that they best match the available data.
    \item By comparing the maximum absolute magnitude that each KN reached in any filter. In cases with sparse data (070809, 150101B) we simply take the maximum observed magnitude.
\end{enumerate}

All three of these methods are flawed - the first may simply measure the brightness of the afterglow, the second does not account for different evolution rates between KNe, and the third does not standardize the filter or the time of measurement. Nonetheless, they represent our best attempts to measure fundamental properties of the KNe.

By metric (i), The KNe (and candidates) associated with GRBs 050709, 070809, 150101B, 160821B and 170817A are all similarly bright; within about half an absolute magnitude of one another. GRBs 060614 and 130603B are both brighter by around two magnitudes. However, GRB 060614 is heavily contaminated by the afterglow according to the models of \citet{Yang15}. The three GRBs with upper limits are all one to two magnitudes fainter.

We also see a very strong correlation between this relative magnitude and \Eiso{} ($p = 4.82\times10^{-3}$ according to the Spearman-r test). However, because they are off axis, GRBs 150101B and 170817A are excluded from this test, so this result is based on just five GRBs. GRBs 050509B, 061201 and 080905A (the upper limit group) are among the lowest $E_{\rm \gamma, iso}$ bursts, along with 160821B.

Metric (ii) shows a less clear picture, with GRBs 070809 (5) and 050709 (4) requiring the second and third highest multiplication factors above the 170817A models, respectively (behind 060614; 6). GRB 130603B (2) requires the second least, higher only than 160821B (0.6), and of course 170817A. There is no trend with the GRB energy.

With metric (iii), 5/7 of the proposed KNe lie in the range of $-15 < M_{\rm abs} < -16$ (070809, 130603B, 150101B, 160821B and 170817A), albeit in a range of filters and measurement times. 060614 is fainter at $-14.35$, but this is likely because the measurement was taken in the I band at around $7$ days after trigger; later and/or in a bluer band than the other six. This hypothesis is reinforced by 060614 being the brightest KN according to metric (i), which standardises the filter and observation time. Contemporaneous J, H or K filter observations were not available. The KN in GRB 050709 is much brighter in absolute magnitude, with $M_K = -17.25$ at $\sim 5$ rest frame days after trigger. GRBs 160821B and 170817A were also both measured in the K-band, at $\sim 4$ and $\sim 3.5$ rest frame days, respectively. GRB 130603B was measured in the H-band at $\sim 7$ days after trigger. Even with the range of measurement times, it seems that the KN in GRB 050709 was unusually bright compared to at least these three other bursts. The three bursts with upper limits show $M_R \leq -13.5$ at $\sim 1.5$ days after trigger (050509B), $M_I \leq -14.5$ at $\sim 3$ days after trigger (061201), and $M_R \leq -13.5$ at $\sim 1.5$ days after trigger (080905A). We also note that GRB 080905A has a very flat evolution, potentially consistent with a KN, at $M_R \approx -14$ at around half a day, though there are only two photometric points to base this on.

The result of these three tests is a mixed picture, and likely reinforces the diversity of KN emission noted by \citet{Gompertz18}, both in terms of brightness and their rate of evolution. The only separation of note is that the \Eiso{} of 130603B and 060614 exceeds the other candidates by at least an order of magnitude (almost two in the case of 060614), while the rest cluster at similar values. They are also observed to be brighter in optical/nIR when the observation times and filters are standardized (metric i)). However, these bursts make an inconvenient pair when searching for clear NS-BH candidates, since 130603B is an NC burst (i.e. a classic SGRB) and 060614 is an unusually luminous EE GRB. Furthermore, both are outshone by the KN in the far less energetic (in \Eiso{} terms) GRB 050709 (an NC burst).

Notably, 050709 was best fitted with an NS-BH merger model \citep{Jin16}, a fact it shares in common with GRB 060614. Their implied ejecta masses are $0.05$~M$_{\odot}$ \citep{Jin16} and $0.1$~M$_{\odot}$ \citep{Yang15} resepectively. 130603B was tested with both NS-NS and NS-BH models, with an inferred ejecta mass in the range $0.03 \leq M_{\odot} \leq 0.08$ \citep{Berger13}. Correspondingly, they are the three KN candidates with the highest estimated ejecta masses. The other four bursts were fitted with NS-NS models. However, because the quoted ejecta masses were derived in different ways, using different models available at the time, direct comparisons should be made with appropriate caution.

\begin{figure*}
\includegraphics[width=\columnwidth]{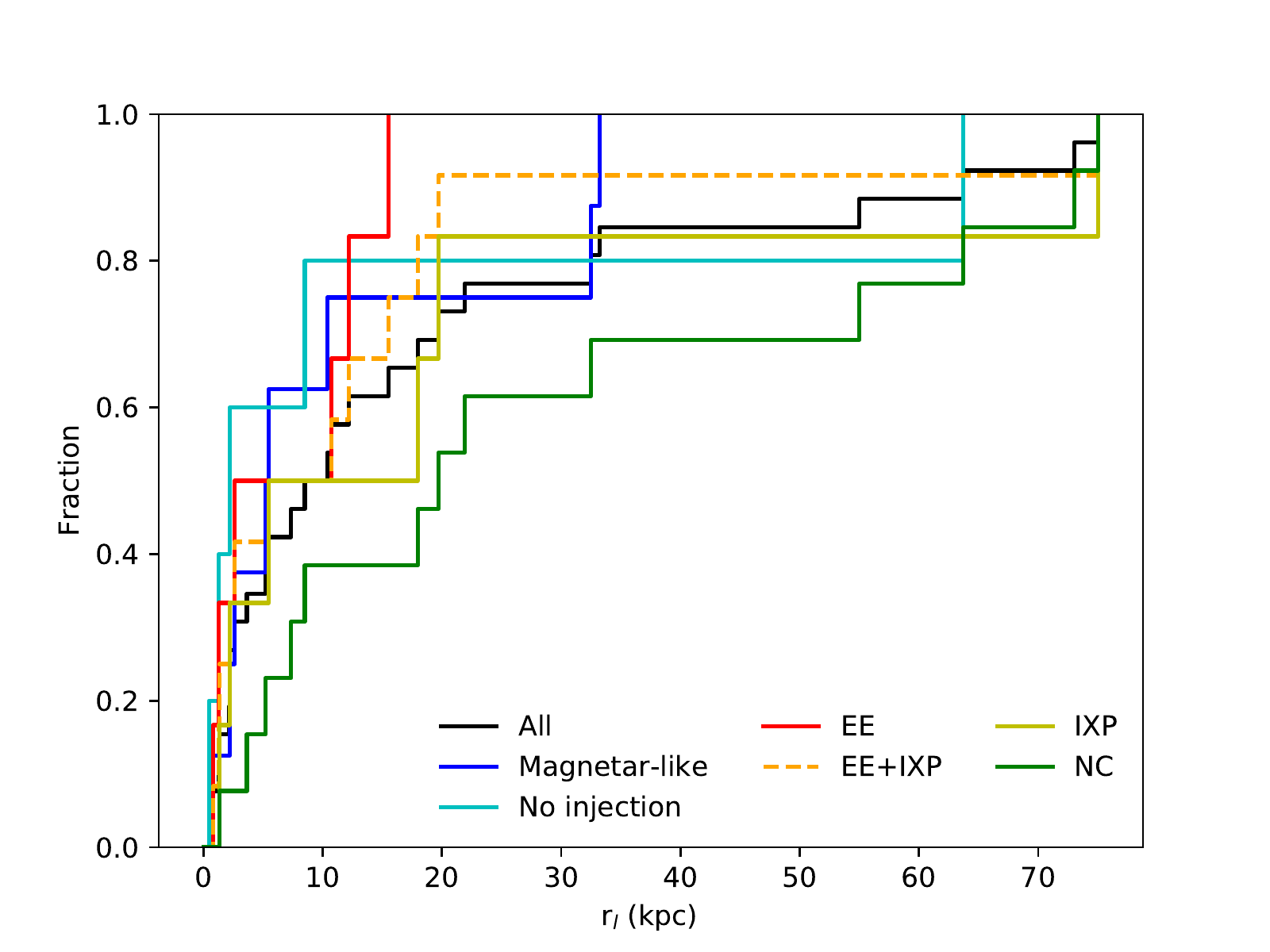}
\includegraphics[width=\columnwidth]{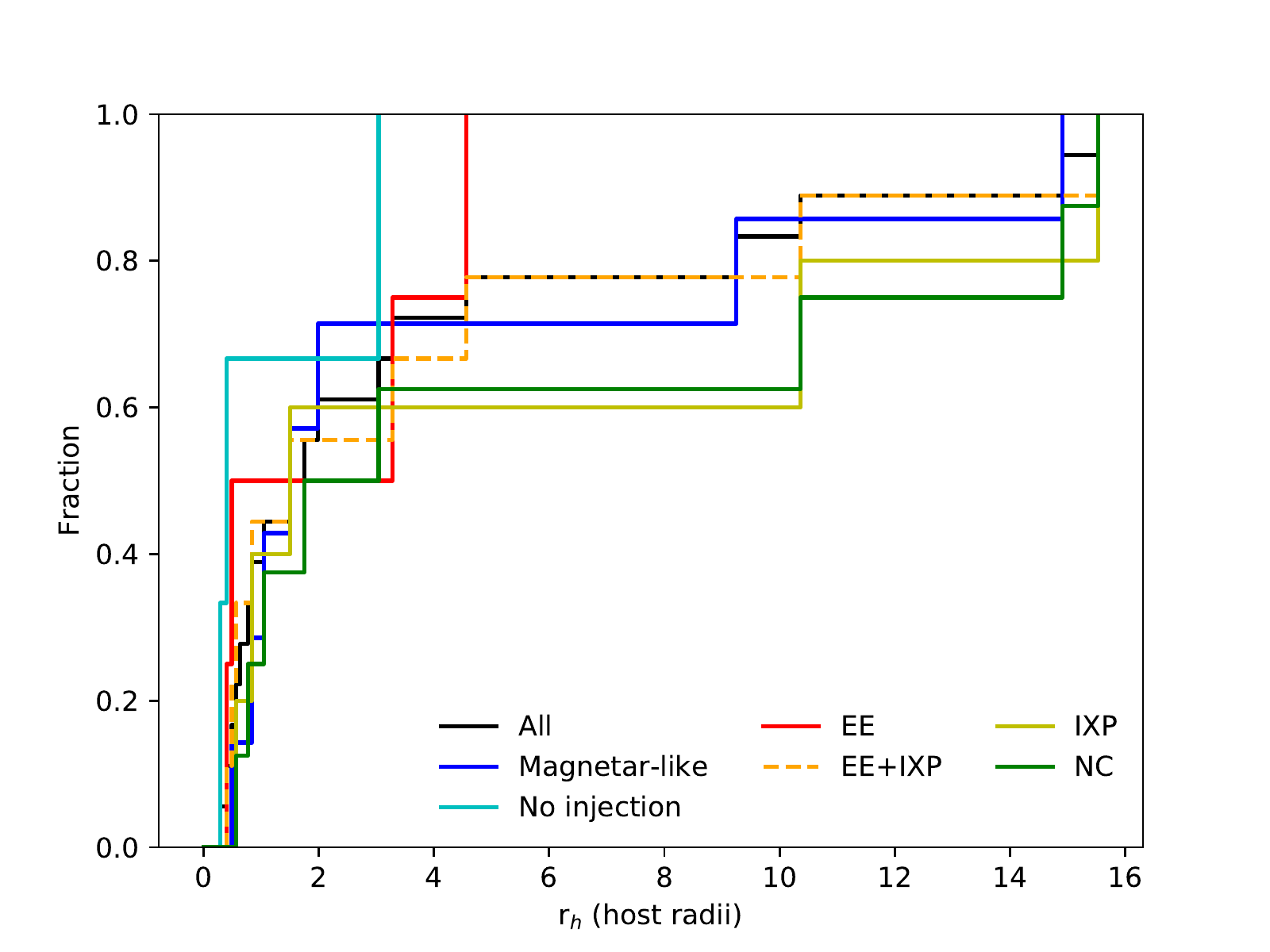}
\caption{The distributions of projected offsets for our sample of SGRBs (black) and assorted sub-samples. Offsets are plotted in absolute terms (in kpc; \emph{left}), and in host-normalised terms (in host effective radii; \emph{right}).}
\label{fig:offsets}
\end{figure*}

\subsection{Host Galaxies}\label{sec:hosts}

Another area in which NS-NS and NS-BH mergers may differ is in their host galaxies. For a given initial separation, a binary system's orbit will decay due to gravitational radiation at a rate proportional to $m_1m_2(m_1+m_2)a^{-4}$ (where $m_1$ and $m_2$ are the constituent masses and $a$ is the initial separation), meaning that (assuming they have the same distribution of initial separations) the lower mass NS-NS binaries will have more time to migrate away from their birth sites before they merge \citep{Belczynski06,Andrews19}. SGRBs associated with NS-BH mergers may therefore be found closer to bright, star-forming regions in their host galaxies, unless their progenitor binaries are formed at systematically greater separations. Such an association has already been suggested in \citet{Troja08}, who found that the five EE GRBs in their sample with measured offsets (not upper limits) had a mean offset of $3.95$~kpc, compared to a mean offset of $31.9$~kpc for the eight non-EE SGRBs with measured offsets (though the latter features many SGRBs that were only localised in X-rays, and hence have large error bars). There is also more mass within a NS-BH binary, such that the impact of momentum conserving kicks will result in a small $\Delta v$, potentially leaving NS-BH binaries closer to their parent galaxies.

However, such an approach is simplistic since the evolution of the binaries to form the double compact object depends on the initial masses and metallicities of the progenitor stars \citep{Eldridge19}. It is quite plausible that the distribution of separation after the formation of the second compact object is very different for NS-NS and NS-BH. Furthermore, the additional mass within the NS-BH systems may keep them bound for larger kick velocities (both natal and binary mass loss related) than for the NS-NS binaries, such that they can survive with larger spatial velocities. Indeed, some studies suggest that the merger times of the more massive NS-BH systems are in fact typically longer than NS-NS binaries \citep[e.g.][]{Toffano19}. Nevertheless, we take an empirical approach to search for  any differences that may exist within the samples.

Host galaxy information for our sample, collected from the literature, is shown in Table~\ref{tab:hosts}. Figure~\ref{fig:offsets} shows the distributions of the offsets of our sample and sub-samples from their host galaxies in terms of their physical offsets in kpc ($r_l$), and in units of measured host galaxy radius ($r_h$). The distributions of $r_l$ in the NC and EE sub-samples are found to be statistically distinct ($p_{\rm KS} = 0.05$; $p_{\rm AD} = 0.04$), though this is not the case for the host normalised offsets, $r_h$. NC bursts are also distinct from non-NC bursts, having a $p_{KS} = 0.04$ and $p_{AD} = 0.01$ probability of their measured $r_l$ being drawn from the same overall distribution. Again, this separation does not hold in $r_h$. The full set of comparisons is available in the Appendix.

\begin{figure}
\includegraphics[width=\columnwidth]{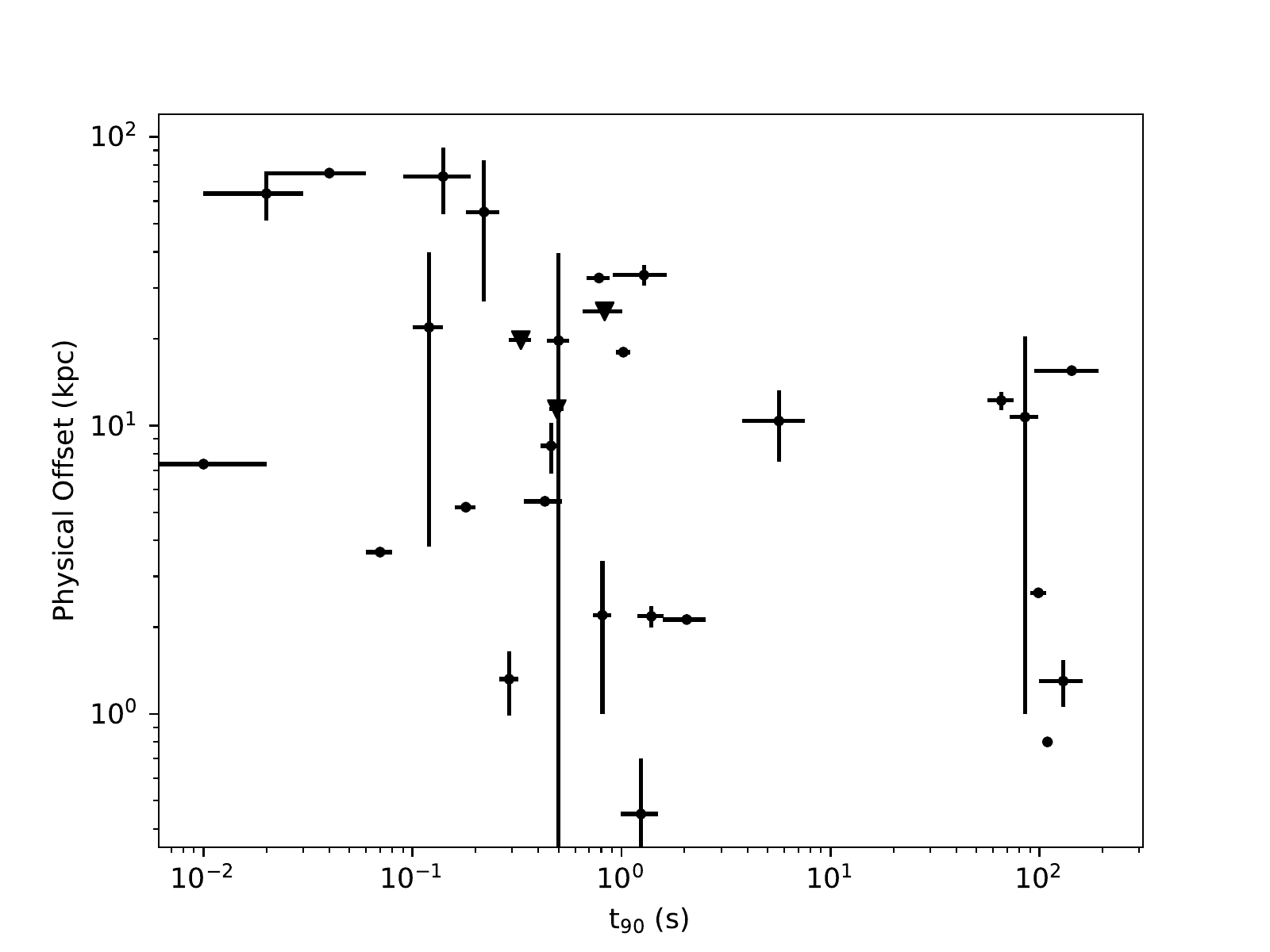}
\caption{The physical offset ($r_l$) of the sample of SGRBs from their putative host galaxies versus t$_{90}$.
\label{fig:rl_vs_t90}}
\end{figure}

We compare the projected physical offsets of the bursts from their host galaxies with their durations ($t_{90}$; Figure~\ref{fig:rl_vs_t90}). To account for the errors in both parameters, we perform 100,000 iterations of the Spearman ranked coefficient test (we choose Spearman-r over Pearson-r because we do not expect the data to follow a normal distribution). On each run, we randomly draw values for each data point from a Gaussian distribution with the value as the mean and the 1$\sigma$ error as the standard deviation. We then take the mean and the standard deviation of our 100,000 Spearman coefficients and p values. We find a correlation coefficient of $-0.38 \pm 0.08$ with $p = 0.09 \pm 0.09$ between $r_l$ and $t_{90}$. When the offsets are normalised by the effective radius ($r_h$) of the host galaxies, we find a Spearman coefficient of $-0.14 \pm 0.05$ and $p = 0.60 \pm 0.12$. This indicates that there is no statistically significant anti-correlation between the durations and host galaxy offsets in our sample.

\begin{figure}
\includegraphics[width=\columnwidth]{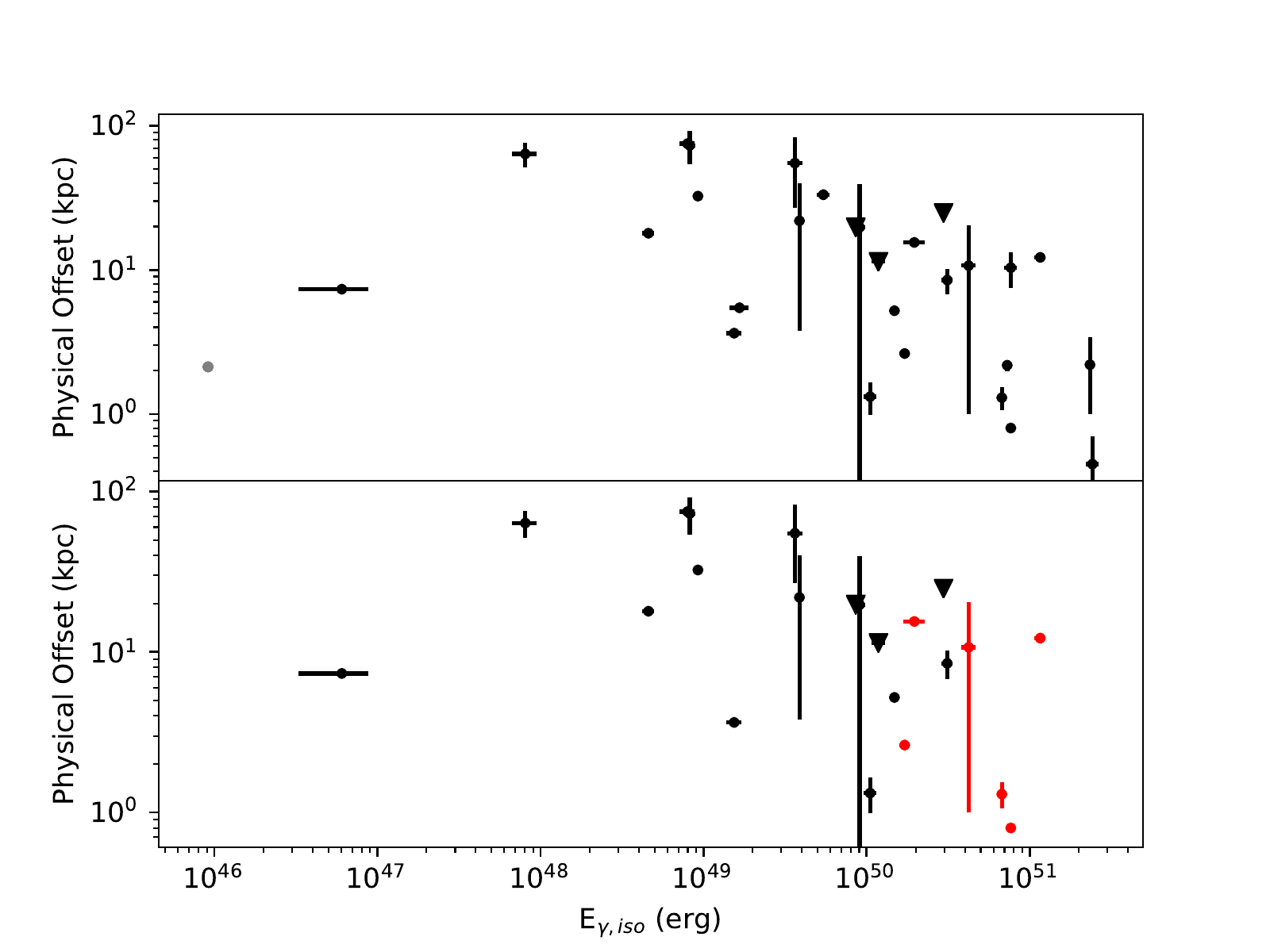}
\caption{The physical offset ($r_l$) of the sample of SGRBs from their putative host galaxies, versus the isotropic energy release of their prompt emission, \Eiso.\newline
\emph{Top}: All bursts in the sample (black). The lone SGRB to the left is 150101B, which is believed to have been viewed off-axis \citep{Troja18}. 170817A, which is known to have been observed off-axis, is also shown in grey.\newline
\emph{Bottom}: NC bursts (black), which have a high probability of being non-collapsars \citep[$f_{\rm NC} \geq 0.5$][]{Bromberg13}, and the EE sub-sample (red).}
\label{fig:rl_vs_Eiso}
\end{figure}

We next investigate the comparison between host galaxy projected offsets and prompt \Eiso. Using the same method as previously, we find a Spearman correlation coefficient of $-0.61 \pm 0.04$ and $p = (1.94 \pm 2.52) \times 10^{-3}$ between \Eiso{} and $r_l$, and a coefficient of $-0.43 \pm 0.03$ with $p = 0.09 \pm 0.03$ between \Eiso{} and $r_h$. Furthermore, the strength of this correlation is clearly diminished by SGRB 150101B, which is alone to the left of the parameter space (Figure~\ref{fig:rl_vs_Eiso}; upper panel). This burst was suggested to have been viewed away from the jet axis \citep{Troja18}, which would result in an under-estimate of its energy release \citep[though weak, on-axis solutions have also been proposed;][]{Fong16,Burns18}. When GRB 150101B is excluded from the Spearman-r test, the coefficients become $-0.67 \pm 0.05$ ($p = [0.84 \pm 1.68] \times 10^{-3}$) for the offsets in kpc, and $-0.57 \pm 0.03$ ($p = 0.02 \pm 0.01$) for offsets normalised by the effective radius of the host. There therefore seems to be an anti-correlation between the energy of a given GRB in our sample and the distance of its afterglow from the putative host galaxy that is statistically significant beyond $3\sigma$. Given that all SGRB afterglows are, by necessity, localised in X-rays before optical detections are made, we do not consider this to be the result of fainter afterglows being harder to detect closer to galaxies, since the X-ray background is not significant. Indeed, several SGRBs in the sample do not have optical detections, and many of the offsets in Table~\ref{tab:hosts} are based on the X-ray position alone (typically those with the largest error bars on the given offset).

Motivated by the marginal statistical distinction in their offset distributions noted earlier, the NC bursts and EE bursts are plotted together in the lower panel of Figure~\ref{fig:rl_vs_Eiso}. The EE bursts exclusively populate the lower-right region of the plot, with high energies and low offsets. This trend is consistent with EE GRBs being a separate population of NS-BH mergers, since high energies and low offsets are the expected characteristics of these mergers. However, we caution that the separation is marginal if $r_l$ and $E_{\rm iso}$ are taken individually, and that the host-normalised locations, $r_h$, are not distinct between the two.

We also note that \citet{Wang18} recently identified an anti-correlation between $r_l$ and \Eiso{} in SGRBs, but only significant to $p = 0.08$ according to the Spearman-r test. Using our method on their data, we find a significance of $p = 0.10\pm0.08$.

One very obvious issue with these results is that we are measuring the 2D projection of a three dimensional offset, and hence the true distance of a given GRB from its host galaxy may be very different when the unknown radial distance is included. However, the solid angle $\Omega = 2\pi (1-\rm cos\theta)$ means that half of the solid angle exists for angles $> 60$ degrees, so that 50 per cent of the binaries will have a real offset that is less than a factor of $2/\sqrt{3} \approx 1.15$ larger than the one we measure. It is therefore far more likely that an offset that appears small in two dimensions is indeed small, rather than masking a much larger offset in the radial direction. Nonetheless, the unknown 3$^{\rm rd}$ dimension will have an effect on our measured correlation, and should be borne in mind when interpreting these results.

Another issue when measuring such a correlation is the uncertainty when assigning a host galaxy that is offset from the afterglow position. Bursts with large $r_l$ are inherently less secure in their host identifications. Furthermore, the `probability of chance alignment' tests that are done to assign hosts in such scenarios \citep[e.g.][]{Berger10,Tunnicliffe14} tend to favour more apparently bright hosts, which would also lower the inferred \Eiso{} in the case of a misidentification \citep{Levan07}. Indeed, four GRBs shown in Table~\ref{tab:hosts} have a measured probability of chance alignment $P_{\rm chance} > 0.05$ \citep{Fong13}. To investigate the importance of this effect, we re-ran our Monte Carlo Spearman-r test with the four high P$_{\rm chance}$ bursts excluded. We find Spearman-r coefficients of $-0.58 \pm 0.07$ ($p = 0.01 \pm 0.02$) and $-0.51 \pm 0.07$ ($p = 0.07 \pm 0.01$) respectively for $r_l$ and $r_h$ versus \Eiso{}, indicating that the significance largely remains. Going a step further, we then re-tested while excluding all bursts with $r_l > 20$~kpc (as well as 150101B). As expected, the correlation is no longer statistically significant; we find Spearman-r coefficients of $-0.44 \pm 0.09$ ($p = 0.10 \pm 0.08$) and $-0.30 \pm 0.06$ ($p = 0.35 \pm 0.10$) respectively for $r_l$ and $r_h$ versus \Eiso{}. The existence of the correlation does therefore depend on whether or not the host galaxies with large offsets have been correctly identified.

\begin{figure}
\includegraphics[width=\columnwidth]{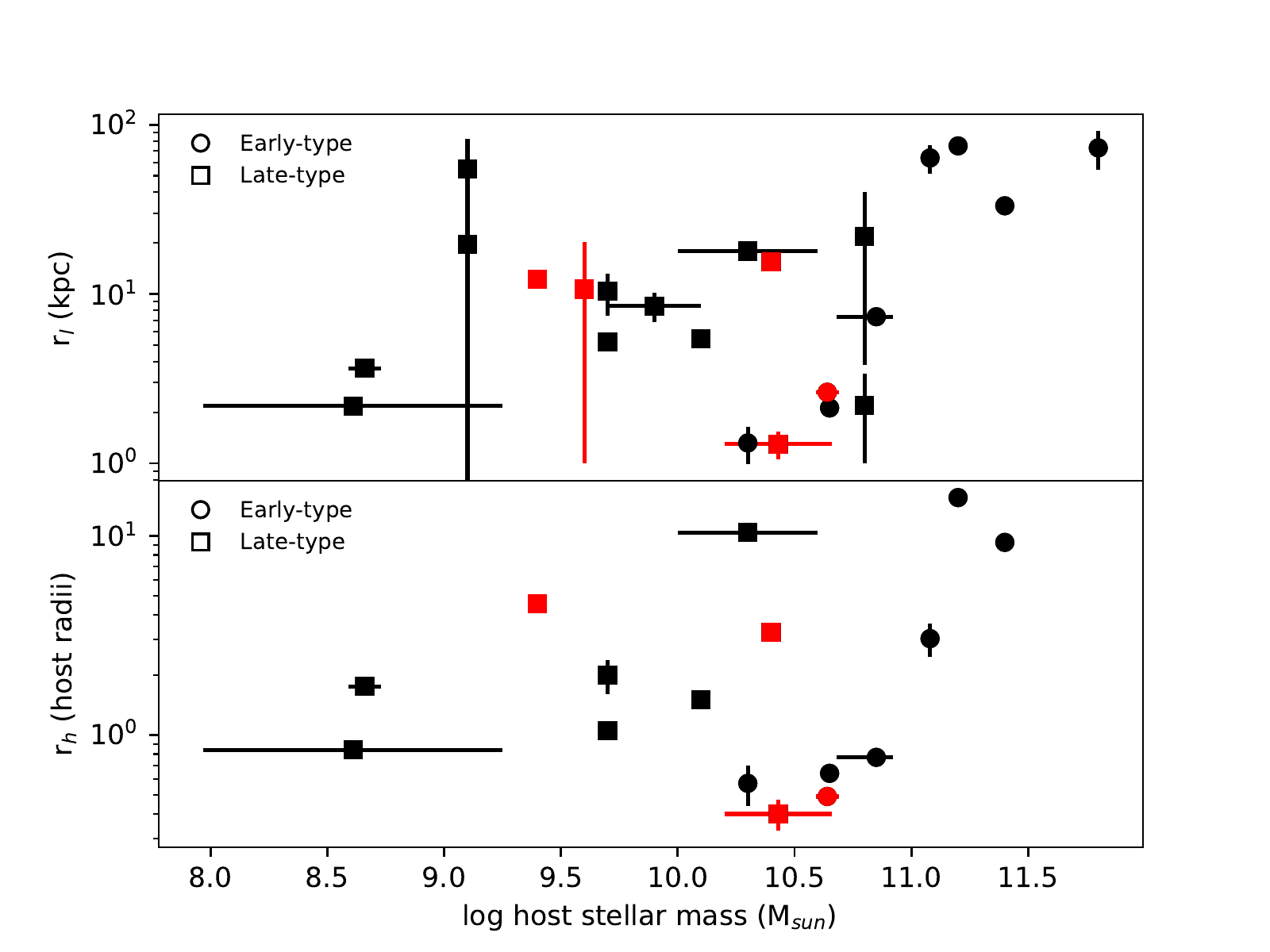}
\caption{\emph{Top}: The physical offset ($r_l$) of our sample vs the stellar mass of their putative host galaxies. Most high offset bursts have been assigned to high mass, early-type galaxies, which are uncommon for the sample as a whole. The EE GRBs are shown in red, with non-EE bursts in black. \newline
\emph{Bottom}: As the top panel, but with the offsets normalised by the radius of the host galaxy.}
\label{fig:galmass}
\end{figure}

Figure~\ref{fig:galmass} illustrates why caution is required; the majority of the high offset bursts appear to reside in massive, early-type galaxies, which is unusual when compared to the rest of the sample. In fact, the physical offset and host stellar mass show a near-significant correlation, with a $p = 0.07$ chance that their alignment is due to random chance according to the Spearman-r test.

To first order, the direction of this correlation is not what would be expected. After formation, a compact object binary orbits in the potential of its host galaxy, or for high velocities may be ejected completely. The stronger galactic potentials in more massive galaxies should hold their binaries closer (at least when normalised by the half light radius to account of the larger physical sizes of more massive galaxies). However, assuming a constant SGRB rate per unit stellar mass, at least some massive elliptical hosts are expected. Furthermore, massive ellipticals also permit longer delay times before merger, and our simple galactic potential argument neglects the fact that these galaxies will grow over time, and may have been less massive when the binary was formed. \citet{Zevin19} showed how the growth history of the host galaxy may enable large host offsets in bursts like 070809 and 090515. \citet{Belczynski06} also found that around a quarter of NS-NS mergers may naturally occur at offsets of several tens of kpc or more from massive elliptical galaxies.

Another possibility is that there is another process which favours the production of SGRBs in massive hosts. That additional process could be the dynamical production of compact object binaries within globular clusters \citep{Grindlay06,Church11}. The specific frequency of globular clusters has been suggested to rise with the galaxy luminosity, such that massive galaxies have proportionally more globular clusters than low mass galaxies \cite{Elmegreen99}. If this is the case we would expect to observe globular cluster created NS-NS and NS-BH binaries preferentially in the most massive galaxies. Since the distribution of globular clusters is much more extended than stars themselves, and rises with galaxy stellar mass \citep{Kartha14} in early-type galaxies (like the high offset hosts in Figure~\ref{fig:galmass}), the larger offsets may be expected. Indeed, a globular cluster origin has been suggested for the largest offset bursts \cite{Church11}.

Recent studies have shown that the rate of compact binary mergers produced in globular clusters is likely to be low. \citet{Belczynski18} find the rate of NS-NS mergers formed dynamically in globular clusters to be $5\times 10^{-5}$~yr$^{-1}$ for all local elliptical galaxies within 100~Mpc$^3$ (compared to $10^{-2}$~yr$^{-1}$ for classical binary mergers in the same volume). Similarly, \citet{Ye20} find the merger rate of NS-NS binaries produced by dynamical interactions in globular clusters to be $0.02$~Gpc$^-3$~yr$^{-1}$ - 5 orders of magnitude below the observed LIGO/Virgo rate. These studies find that globular clusters are not significant contributors to the compact object merger rate, and combined with our analysis may indicate that a non-negligible fraction of SGRB hosts are misidentified - an important result in itself.

The high offset `hostless' GRBs have been thoroughly investigated, and even with deep Hubble Space Telescope images, no underlying hosts have yet been discovered \citep[e.g.][]{Fong10,Fong13b}. We attempt to quantify how sensitive our correlation is to the misidentification of the host galaxy by repeating our Monte Carlo Spearman-r test while randomly drawing n-1, n-2 and n-3 galaxies from our sample (where n is the total number of SGRBs with a measured $r_l$). We find that even when three bursts are excluded, the measured correlation between $r_l$ and \Eiso{} falls short of the $p = 0.05$ significance threshold in just 335 of our 100,000 runs ($0.34$ per cent), indicating that the correlation is not heavily relying on any one (or even three) burst(s). We also record which SGRBs were removed when the correlation falls short of the significance threshold. The largest contributors to failed runs were GRBs 090426, 050509B, 090515 and 060502B.

GRB 090426 may in fact be an interloping LGRB; \citet{Bromberg13} assign it a probability of being a non-collapsar (i.e. an SGRB) of $f_{NC} = 0.10^{+0.15}_{-0.06}$ \citep[see however][]{Thoene11}. This fact highlights another potential confounding factor for our measured correlation - namely that any interloping LGRBs will naturally be more energetic and lie closer to their hosts \citep[see][]{Fruchter06} than the SGRB population, thus biasing our correlation at low offsets/high energies. To test this, we measure the correlation for our sub-sample of NC bursts. We find that the correlation still holds, with a Spearman coefficient of $-0.60 \pm 0.07$ and $p = 0.046 \pm 0.035$ as per our previous method. Because $f_{NC}$ is calculated from $t_{90}$ and a single power law fit to the prompt emission spectral slope, EE GRBs are naturally excluded. This means that even when our sample is stripped to high probability pure SGRBs, the anti-correlation between physical offsets from host galaxies and prompt emission energies is still observed, though given the aforementioned caveats, its explanation is far from simple. At very least, we can confirm that it is not entirely due to interloping LGRBs biasing the high energy -- low offset end.

Two other SGRBs of note are 050509B and 061201. 050509B is the burst whose exclusion was the second highest contributor to Monte Carlo runs that fell short of being statistically significant. This SGRB has been assigned to a large early-type galaxy at $z = 0.225$ \citep{Castro-Tirado05,Hjorth05b}. However, this host is a significant outlier in the distribution of host galaxy sizes in our sample. While the host of SGRB\,050509B is measured to have an effective radius of $r_e = 20.98$~kpc \citep{Fong10}, the rest of the sample have a mean effective radius of $\bar{r}_e = 3.77 \pm 1.95$~kpc (excluding 050509B). Coupled with the fact that 050509B has the lowest \Eiso{} in the sample (excluding the off-axis 150101B), this may suggest that the host galaxy has been misidentified in this case, although its presence in a cD galaxy of a merging cluster also suggests that the probability of chance alignment is genuinely small. The host of GRB 061201 is also uncertain \citep{Stratta07,Berger09,Fong10}. It may in fact be associated with a faint galaxy at $z \gtrsim 1$. However, even if this is the case, its offset would be lower \citep[$14.47\pm 0.24$~kpc;][]{Fong13b}, and the greater implied energy of \Eiso{} $\sim 8\times10^{50}$~erg means that it would still follow our observed trend.

\section{Conclusions}\label{sec:conclusions}

We investigated the complete sample of SGRBs with redshift for any evidence of a dichotomy that would indicate that both the NS-BH and NS-NS formation channels operate. The inhomogeneity of the available data makes classifications and comparisons difficult in general, but one group within the sample, the EE GRBs, do show several characteristics that tentatively support the idea that they are a distinct phenomenon.

First, the durations of EE bursts, as measured by \emph{Swift}-BAT, are statistically distinct when compared to the non-EE sub-sample. This is by definition of the sub-sample, but is nonetheless a distinguishing property. \Eiso{} is also distinct, but not when the analysis is either limited to the first two seconds of EE (as an analogue of regular SGRBs), or when all data ($\gamma$-rays + X-rays) up to the average duration of EE are included.

Second, EE bursts have marked differences in their durations when measured in the 15 -- 50 keV bandpass compared to the 50 -- 100 keV bandpass, a trait which is not shared by either SGRBs or LGRBs, whose durations in these bandpasses largely track a 1:1 ratio. This may indicate an additional emission process, for example fallback material due to tidal stripping of an unequal mass binary.

Third, an AD test reveals that the physical offsets of EE GRBs and SGRBs with $f_{\rm NC} \geq 0.5$ \citep{Bromberg13} from their host galaxies have a $p = 0.04$ probability of being drawn from the same populationIn this context, the NC bursts are a useful sub-sample that takes some extra steps in mitigating against interlopers from the collapsar distribution, and hence are likely to be purer than the general SGRB sample. The trade-off is that some atypical SGRBs are omitted, which restricts the sample size and may potentially introduce unseen bias if the omitted fraction is significantly large and/or significantly atypical. EE GRBs are found at systematically lower offsets than SGRBs, a property which, when coupled with their greater energy release, agrees with the expectations of a formation channel involving a higher mass binary with a shorter merger time. However, the statistical separation between the two sub-samples does not persist when normalised for the effective radius of the host galaxy.

One major implication of the possibility that EE GRBs are NS-BH mergers is the fate of the magnetar model; if EE GRBs are indeed shown to be NS-BH mergers, they cannot produce magnetars, and so the model can effectively be ruled out for SGRBs in general. This is because both EE GRBs and pure SGRBs show magnetar-like plateaus \citep{Gompertz13}, and due to their similarity it is unlikely that they are due to two distinct mechanisms. We do not find any statistically significant distinctions between GRBs whose afterglows are well fitted with the magnetar model versus those that are well fitted with a simple power law (indicating no energy injection), except that the latter are at higher redshifts according to the AD test. In the absence of other appreciable differences, the best explanation for this is likely to be that magnetar-like injection plateaus are harder to identify at higher redshifts.

A confounding factor is the internal X-ray plateau bursts, which show similar emission features to EE GRBs, but at energies lower than the BAT bandpass. Several of the lowest offset non-EE SGRBs are indeed bursts with internal X-ray plateaus, but conversely, so is the highest offset burst (090515). Internal X-ray plateau bursts are also not distinct in energy or duration from regular SGRBs. A more detailed investigation specifically into their nature may shed light on whether EE is a distinct class, or the extreme end of a single distribution.

Finally, we find a statistically significant anti-correlation between the physical offset of a given SGRB from its host galaxy and its prompt emission energy, \Eiso. This correlation holds for the sample as a whole, as well as for our NC sub-sample, which is filtered to remove any interloping LGRBs \citep{Bromberg13}, and naturally excludes EE. Its interpretation is complicated by the unknown offset in the radial direction. Based on solid angle arguments, the observed offset is likely to be close to the true offset in most cases, but the impact of this uncertainty is nonetheless unknown. If the correlation is real, it is robust against the removal at random of up to three GRBs from the sample, but is somewhat sensitive to the correct identification of the host galaxies of high offset bursts.

Many of the highest offset GRBs in our sample are associated with massive elliptical host galaxies. These galaxies are unusual when compared to the typical SGRB host, raising concerns about whether they have been correctly identified. More massive galaxies will have the highest escape velocities, but also afford the longest delay times before merger, and may have been significantly less massive when the binary formed \citep{Zevin19}. It is unlikely that natal kick velocities play a significant role because the binary will complete many orbits of its host between formation and merger.

For our observed correlation to be invalidated, more than half of the SGRBs with offsets of more than 20~kpc from their host galaxies would have to be incorrectly identified - an important result in itself. It is also possible that there is a population of SGRBs in globular clusters, the number and radial extent of which does correlate with galaxy mass. However, the expected merger rate via the globular cluster channel is very low \citep{Belczynski18,Ye20}.

\section*{Acknowledgements}

We thank the anonymous referee for a careful reading of the manuscript, and thoughtful suggestions that helped to crystallise the discussion in this paper.

BG,  AJL \& NRT have received funding from the European Research Council (ERC) under the European Union's Horizon 2020 research and innovation programme (grant agreement no 725246, TEDE, PI Levan). AJL acknowledges support from STFC via grant ST/P000495/1. We
also gratefully acknowledge the use of enhanced \emph{Swift} data products from the UKSSDC at the University of Leicester.



\bibliographystyle{aasjournal}
\bibliography{ref}



\appendix

\begin{longtable*}{lrccccccc}
\hline\hline
{\bf E$_{\rm iso}$} & \vline & & & & AD & & & \\
 & \vline & Full & EE & EE+IXP & IXP & NC & M & IF \\
\hline
 & Full \vline & ----- & \textcolor{black}{$4.37 \times 10^{-3}$} & \textcolor{red}{$> 0.25$} & \textcolor{red}{$0.06$} & \textcolor{blue}{$< 10^{-3}$} & \textcolor{red}{$> 0.25$} & \textcolor{red}{0.05} \\
 & EE \vline & \textcolor{blue}{$1.33 \times 10^{-3}$} & ----- & ----- & \textcolor{black}{$3.06 \times 10^{-3}$} & \textcolor{blue}{$< 10^{-3}$} & \textcolor{red}{$0.14$} & \textcolor{red}{$> 0.25$} \\
 & EE+IXP \vline & \textcolor{red}{$0.81$} & ----- & ----- & ----- & \textcolor{red}{$0.06$} & \textcolor{red}{$> 0.25$} & \textcolor{red}{$> 0.25$} \\
KS & IXP \vline & \textcolor{red}{$0.10$} & \textcolor{blue}{$1.40 \times 10^{-3}$} & ----- & ----- & \textcolor{red}{$> 0.25$} & \textcolor{red}{$0.09$} & \textcolor{black}{$0.03$} \\
 & NC \vline & \textcolor{blue}{$8.86 \times 10^{-4}$} & \textcolor{blue}{$1.54 \times 10^{-4}$} & \textcolor{red}{$0.13$} & \textcolor{red}{$0.93$} & ----- & \textcolor{black}{$0.03$} & \textcolor{black}{$3.99 \times 10^{-3}$} \\
 & M \vline & \textcolor{red}{0.51} & \textcolor{red}{$0.19$} & \textcolor{red}{$0.95$} & \textcolor{red}{$0.14$} & \textcolor{red}{$0.13$} & ----- & \textcolor{red}{$> 0.25$} \\
 & IF \vline & \textcolor{red}{0.07} & \textcolor{red}{$0.43$} & \textcolor{red}{$0.65$} & \textcolor{black}{$0.04$} & \textcolor{black}{$0.04$} & \textcolor{red}{$0.79$} & ----- \\
\hline\hline

{\bf E$_{\rm iso}$} & ($1^{\rm st}$ 2s only) \vline & Full & IXP & NC & M & IF \\
\hline
KS & EE \vline & \textcolor{red}{$0.76$} & \textcolor{red}{$0.79$} & \textcolor{red}{$0.95$} & \textcolor{red}{$0.99$} & \textcolor{red}{$0.15$} \\
AD & EE \vline & \textcolor{red}{$> 0.25$} & \textcolor{red}{$> 0.25$} & \textcolor{red}{$> 0.25$} & \textcolor{red}{$> 0.25$} & \textcolor{red}{$> 0.25$} \\
\hline\hline

{\bf t$_{\rm 90}$} & \vline & & & & AD & & & \\
 & \vline & Full & EE & EE+IXP & IXP & NC & M & IF \\
\hline
 & Full \vline & ----- & \textcolor{blue}{$< 10^{-3}$} & \textcolor{black}{$5.01 \times 10^{-3}$} & \textcolor{red}{$> 0.25$} & \textcolor{blue}{$< 10^{-3}$} & \textcolor{red}{$0.05$} & \textcolor{red}{$> 0.25$} \\
 & EE \vline & \textcolor{blue}{$1.58 \times 10^{-7}$} & ----- & ----- & \textcolor{blue}{$< 10^{-3}$} & \textcolor{blue}{$< 10^{-3}$} & \textcolor{black}{$9.93 \times 10^{-3}$} & \textcolor{blue}{$2.34 \times 10^{-3}$} \\
 & EE+IXP \vline & \textcolor{black}{$0.04$} & ----- & ----- & ----- & \textcolor{blue}{$< 10^{-3}$} & \textcolor{red}{$> 0.25$} & \textcolor{red}{$> 0.25$} \\
KS & IXP \vline & \textcolor{red}{$0.42$} & \textcolor{blue}{$1.75 \times 10^{-4}$} & ----- & ----- & \textcolor{red}{$0.10$} & \textcolor{red}{$0.09$} & \textcolor{red}{$> 0.25$} \\
 & NC \vline & \textcolor{blue}{$1.09 \times 10^{-6}$} & \textcolor{blue}{$1.28 \times 10^{-6}$} & \textcolor{blue}{$1.95 \times 10^{-3}$} & \textcolor{red}{$0.33$} & ----- & \textcolor{blue}{$< 10^{-3}$} & \textcolor{red}{$0.09$} \\
 & M \vline & \textcolor{red}{$0.08$} & \textcolor{black}{$0.02$} & \textcolor{red}{$1.00$} & \textcolor{red}{$0.14$} & \textcolor{black}{$7.02 \times 10^{-3}$} & ----- & \textcolor{red}{$> 0.25$} \\
 & IF \vline & \textcolor{red}{$0.99$} & \textcolor{blue}{$8.23 \times 10^{-4}$} & \textcolor{red}{$0.40$} & \textcolor{red}{$0.96$} & \textcolor{red}{$0.33$} & \textcolor{red}{$0.42$} & ----- \\
\hline\hline

{\bf z} & \vline & & & & AD & & & \\
 & \vline & Full & EE & EE+IXP & IXP & NC & M & IF \\
\hline
 & Full \vline & ----- & \textcolor{red}{$0.23$} & \textcolor{red}{$> 0.25$} & \textcolor{red}{$> 0.25$} & \textcolor{red}{$> 0.25$} & \textcolor{red}{$0.10$} & \textcolor{red}{$0.06$} \\
 & EE \vline & \textcolor{red}{$0.08$} & ----- & ----- & \textcolor{red}{$> 0.25$} & \textcolor{red}{$> 0.25$} & \textcolor{red}{$> 0.25$} & \textcolor{black}{$0.02$} \\
 & EE+IXP \vline & \textcolor{red}{$0.26$} & ----- & ----- & ----- & \textcolor{red}{$> 0.25$} & \textcolor{red}{$> 0.25$} & \textcolor{red}{$0.06$} \\
KS & IXP \vline & \textcolor{red}{$0.83$} & \textcolor{red}{$0.17$} & ----- & ----- & \textcolor{red}{$> 0.25$} & \textcolor{red}{$> 0.25$} & \textcolor{red}{$> 0.25$} \\
 & NC \vline & \textcolor{red}{$0.97$} & \textcolor{red}{$0.27$} & \textcolor{red}{$0.67$} & \textcolor{red}{$0.88$} & ----- & \textcolor{red}{$> 0.25$} & \textcolor{red}{$0.18$} \\
 & M \vline & \textcolor{red}{$0.13$} & \textcolor{red}{$0.89$} & \textcolor{red}{$0.58$} & \textcolor{red}{$0.39$} & \textcolor{red}{$0.45$} & ----- & \textcolor{black}{$0.05$} \\
 & IF \vline & \textcolor{red}{$0.11$} & \textcolor{black}{$0.04$} & \textcolor{red}{$0.18$} & \textcolor{red}{$0.39$} & \textcolor{red}{$0.29$} & \textcolor{red}{$0.17$} & ----- \\
\hline\hline

{\bf r$_l$} & \vline & & & & AD & & & \\
 & \vline & Full & EE & EE+IXP & IXP & NC & M & IF \\
\hline
 & Full \vline & ----- & \textcolor{red}{$> 0.25$} & \textcolor{red}{$> 0.25$} & \textcolor{red}{$> 0.25$} & \textcolor{black}{$0.01$} & \textcolor{red}{$> 0.25$} & \textcolor{red}{$0.20$} \\
 & EE \vline & \textcolor{red}{$0.24$} & ----- & ----- & \textcolor{red}{$> 0.25$} & \textcolor{black}{$0.04$} & \textcolor{red}{$> 0.25$} & \textcolor{red}{$> 0.25$} \\
 & EE+IXP \vline & \textcolor{red}{$0.35$} & ----- & ----- & ----- & \textcolor{red}{$0.09$} & \textcolor{red}{$> 0.25$} & \textcolor{red}{$> 0.25$} \\
KS & IXP \vline & \textcolor{red}{$0.97$} & \textcolor{red}{$0.47$} & ----- & ----- & \textcolor{red}{$> 0.25$} & \textcolor{red}{$> 0.25$} & \textcolor{red}{$> 0.25$} \\
 & NC \vline & \textcolor{black}{$0.04$} & \textcolor{red}{$0.05$} & \textcolor{red}{$0.23$} & \textcolor{red}{$0.77$} & ----- & \textcolor{red}{$0.16$} & \textcolor{red}{$0.09$} \\
 & M \vline & \textcolor{red}{$0.58$} & \textcolor{red}{$0.95$} & \textcolor{red}{$0.90$} & \textcolor{red}{$0.95$} & \textcolor{red}{$0.32$} & ----- & \textcolor{red}{$> 0.25$} \\
 & IF \vline & \textcolor{red}{$0.41$} & \textcolor{red}{$0.90$} & \textcolor{red}{$0.84$} & \textcolor{red}{$0.69$} & \textcolor{red}{$0.21$} & \textcolor{red}{$0.77$} & ----- \\
\hline\hline

{\bf r$_h$} & \vline & & & & AD & & & \\
 & \vline & Full & EE & EE+IXP & IXP & NC & M & IF \\
\hline
 & Full \vline & ----- & \textcolor{red}{$> 0.25$} & \textcolor{red}{$> 0.25$} & \textcolor{red}{$> 0.25$} & \textcolor{red}{$0.21$} & \textcolor{red}{$> 0.25$} & \textcolor{black}{$0.05$} \\
 & EE \vline & \textcolor{red}{$0.54$} & ----- & ----- & \textcolor{red}{$> 0.25$} & \textcolor{red}{$0.24$} & \textcolor{red}{$> 0.25$} & \textcolor{red}{$> 0.25$} \\
 & EE+IXP \vline & \textcolor{red}{$0.99$} & ----- & ----- & ----- & \textcolor{red}{$> 0.25$} & \textcolor{red}{$> 0.25$} & \textcolor{red}{$0.11$} \\
KS & IXP \vline & \textcolor{red}{$0.74$} & \textcolor{red}{$0.56$} & ----- & ----- & \textcolor{red}{$> 0.25$} & \textcolor{red}{$> 0.25$} & \textcolor{red}{$0.17$} \\
 & NC \vline & \textcolor{red}{$0.49$} & \textcolor{red}{$0.51$} & \textcolor{red}{$0.94$} & \textcolor{red}{$0.99$} & ----- & \textcolor{red}{$> 0.25$} & \textcolor{red}{$0.09$} \\
 & M \vline & \textcolor{red}{$0.67$} & \textcolor{red}{$0.78$} & \textcolor{red}{$0.99$} & \textcolor{red}{$0.96$} & \textcolor{red}{$0.95$} & ----- & \textcolor{red}{$0.15$} \\
 & IF \vline & \textcolor{red}{$0.16$} & \textcolor{red}{$0.66$} & \textcolor{red}{$0.45$} & \textcolor{red}{$0.29$} & \textcolor{red}{$0.19$} & \textcolor{red}{$0.25$} & ----- \\
\hline\hline
\end{longtable*}
\noindent \textbf{A1:} Summary of our KS and AD tests used throughout the paper. `Full' refers to all GRBs not included in the comparison category. \textcolor{blue}{Blue} cells indicate $p \leq 0.003$ ($3\sigma$ separation), \textcolor{black}{black} cells indicate $0.003 < p \leq 0.05$ ($2\sigma$ separation), and \textcolor{red}{red} cells indicate that the two sub-samples are consistent with being drawn from a single distribution. Different colours of a boundary number (e.g. $p = 0.05$) indicate whether the value was rounded up (black) or down (red).

\end{document}